\documentclass[12pt]{imsart}
\usepackage{amsthm,amsmath}
\usepackage[colorlinks,citecolor=blue,urlcolor=blue]{hyperref}

\usepackage{amsmath,amsfonts,amssymb,amsthm}

\usepackage{enumitem}

\RequirePackage{amsthm,amsmath,amsfonts,amssymb}
\RequirePackage[numbers]{natbib}
\RequirePackage[colorlinks,citecolor=blue,urlcolor=blue]{hyperref}
\RequirePackage{graphicx}
\usepackage{lineno,hyperref}

\startlocaldefs
\modulolinenumbers[5]

\usepackage[utf8]{inputenc}
\usepackage[english]{babel}

\usepackage[margin=1in]{geometry}
\usepackage{bm}
\usepackage{graphicx}
\usepackage{amssymb,amsmath,amsthm,mathrsfs}
\usepackage{epstopdf}
\usepackage{algorithm}
\usepackage{amsfonts,delarray}
\usepackage{algorithm,algcompatible,amsmath}
\usepackage{rotating,color}
\usepackage{subfigure}
\usepackage{multirow}
\usepackage{titlesec,textcase}
\usepackage{longtable}
\usepackage{bbm}
\usepackage{rotating}

\newtheorem{Theorem}{Theorem}[section]
\newtheorem{Lemma}[Theorem]{Lemma}

\newtheorem{Proposition}[Theorem]{Proposition}

\newtheorem{Definition}[Theorem]{Definition}

\theoremstyle{definition}
\newtheorem{Example}{Example}[section]

\RequirePackage[normalem]{ulem} %DIF PREAMBLE
\definecolor{rp}{RGB}{83,54,106}

\def\boxit#1{\vbox{\hrule\hbox{\vrule\kern6pt\vbox{\kern6pt#1\kern6pt}\kern6pt\vrule}\hrule}}

\endlocaldefs

\allowdisplaybreaks

\begin{document}
\begin{frontmatter}
\title{Testing common invariant subspace of multilayer networks}

\runtitle{Testing common invariant subspace }
%\thankstext{T1}{A sample additional note to the title.}
\runauthor{ }
\begin{aug}

\author[A]{\fnms{Mingao} \snm{Yuan}\ead[label=e1]{mingao.yuan@ndsu.edu}}\and
\author[B]{\fnms{Qianqian} \snm{Yao}\ead[label=e2]{qianqian.yao@ndsu.edu}}
%%%%%%%%%%%%%%%%%%%%%%%%%%%%%%%%%%%%%%%%%%%%%%
%% Addresses                                %%
%%%%%%%%%%%%%%%%%%%%%%%%%%%%%%%%%%%%%%%%%%%%%%
%\address[C]{School of Mathematics and Statistics,
%North China University of Water Resources and Electric Power,
%\printead{e3}}

%\address[B]{Department of Statistics,
%North Dakota State University,
%\printead{e2}}

\address[A]{Department of Statistics ,
North Dakota State University, Fargo, ND, USA
\printead{e1}}
\address[B]{Department of Statistics,
North Dakota State University, Fargo, ND, USA
\printead{e2}}

\end{aug}

\begin{abstract}
 Graph (or network) is a mathematical structure that has been widely used to model relational data. As real-world systems get more complex, multilayer (or multiple) networks are employed to represent diverse patterns of relationships among the objects in the systems. One active research problem in multilayer networks analysis is to study the common invariant subspace of the networks, because such common invariant subspace could capture the fundamental structural patterns and interactions across all layers. Many methods have been proposed to estimate the common invariant subspace.  However, whether real-world multilayer networks share the same common subspace remains unknown. In this paper, we first attempt to answer this question by means of hypothesis testing. The null hypothesis states that the multilayer networks share the same subspace, and under the alternative hypothesis, there exist at least two networks that do not have the same subspace. We propose a Weighted Degree Difference Test, derive the limiting distribution of the test statistic and provide an analytical analysis of the power. Simulation study shows that the proposed test has satisfactory performance, and a real data application is provided.
\end{abstract}

\begin{keyword}[class=MSC2020]
\kwd[]{60K35}
\kwd[;  ]{05C80}
\end{keyword}

\begin{keyword}
\kwd{multilayer graphs}
\kwd{common subspace}
\kwd{hypothesis test}
\end{keyword}

\end{frontmatter}

\section{Introduction}
\label{S:1}

Graphs (or Networks) are a widely used data structure and a common language to model connected data for describing complex systems. Essentially, a graph ( or a network) consists of a collection of nodes to represent objects and a set of edges to represent interactions between pairs of these objects. 
One of the popular mathematical expressions of a graph is by an adjacency matrix, where the rows and columns correspond to the graph's nodes, and numerical values indicate the presence of edges between nodes.  Graph data can be found in a broad spectrum of application domains. For example, graphs can be applied to model social networks, where nodes represent individuals or entities, and edges typically denote friendships, collaborations, interactions, or other social ties \cite{Li21}. Graph models are also employed to model molecules in quantum chemistry, catalyst discovery, drug discovery, etc. to predict the properties of molecules.  The atoms in molecules are modeled as nodes and the bonds between two atoms are modeled as edges \cite{Gilmer17, SS23, DTM20, Yao22}.

Due to its widespread applications, graph data mining has gained tremendous popularity in the past decades.  Some typical tasks of graph data mining include link prediction, node classification, community detection, testing community structure and so on \cite{A17,AV14,BS16,GMZZ18,JKL21,L16,YYS22,ZLZ12,YLFS20,YS22,HWX16,SBL13,Xia21,NLL23,CS10,YLFS20,YZZ24}.   Most of the existing methods are developed for single network.

As research on complex systems has progressed, multilayer (or multiple) networks have attracted a lot of attentions\cite{D17,DP15,LLK20,CLM22,Kivela14}. Multilayer networks can be used to model multiple types of interactions that a single network cannot. Multilayer networks consist of a group of networks, with each network representing different types of interactions.  Each layer operates under its own rules and dynamics, reflecting the diversity of interactions within the system. For instance, in multilayer social networks, one layer might represent personal friendships, another professional collaboration, and a third might capture shared interests or activities \cite{OKH14}. Application of multilayer networks in biomedicine is discussed thoroughly in \cite{HK20}. An overview of multilayer network analysis and its application to epidemiological research questions has been proposed in \cite{KRSV20}.  There are several types of multilayer networks in literature. In this paper, we focus on multilayer networks (or multilayerx networks) defined on the same set of nodes, with edges only connecting nodes within the same layer \cite{D17,DP15,LLK20,CLM22,Arroyo21, PW24, Pensky24}. 

One quite active research problem in multilayer networks analysis is to study the common invariant subspace of the networks\cite{AP04, ZT22, Wang18, Arroyo21, PW24, Pensky24}. A common invariant subspace of multilayer networks refers to a shared latent space that captures the fundamental structural patterns and interactions across all layers. Such subspace serves as a unified representation that harmonizes the heterogeneous data, allowing for integrated analysis and learning. By embedding each layer of network into this common subspace, one can uncover the underlying principles governing the interactions, which remain consistent despite the diversity of the individual networks. Such subspace can further improve accuracy, robustness and adaptability of machine learning models \cite{Chen24}. In \cite{Arroyo21}, the authors  proposed a spectral algorithm to estimate the common invariant subspace of multilayer networks. \cite{PW24, Pensky24} studied the problem of clustering multilayer networks into groups such that each group shares the same common invariant subspace.
 
Given real-world multilayer networks, a natural question is whether the networks share the same common subspace. 
In this paper, we initiate the study of this problem and formulate it as a hypothesis testing problem as follows. Under the null hypothesis, the multilayer networks share the same common subspace, while under the alternative hypothesis, the multilayer networks do not have the same common subspace. We propose a Weighted Degree Difference Test, which is the first test for testing common invariant subspace of multilayer networks. Under the null hypothesis, we prove that the test statistic converges in distribution to the standard normal distribution as the number of nodes goes to infinity. We also provide an analytical characterization of the asymptotic power of the test. Simulation study shows the proposed test has good performance, and a real data application is presented.

The paper is organized as follows. In Section \ref{mainRe}, we formally introduce the hypotheses and the proposed test. In Section \ref{SimRD}, we report simulation results and a real data application.  Section \ref{mProf} presents the proofs.

\vskip 5mm

\noindent
{\bf Notation:} We adopt the  Bachmann–Landau notation throughout this paper. Let $a_n$  and $b_n$ be two positive sequences. Denote $a_n=\Theta(b_n)$ if $c_1b_n\leq a_n\leq c_2 b_n$ for some positive constants $c_1,c_2$. Denote  $a_n=\omega(b_n)$ if $\lim_{n\rightarrow\infty}\frac{a_n}{b_n}=\infty$. Denote $a_n=O(b_n)$ if $a_n\leq cb_n$ for some positive constants $c$. Denote $a_n=o(b_n)$ if $\lim_{n\rightarrow\infty}\frac{a_n}{b_n}=0$. Let $X_n,X$ be random variables. Denote $X_n=O_P(a_n)$ if $\frac{X_n}{a_n}$ is bounded in probability. Denote $X_n=o_P(a_n)$ if $\frac{X_n}{a_n}$ converges to zero in probability as $n$ goes to infinity. Let $\mathbb{E}[X]$ and $Var(X)$ denote the expectation and variance of a random variable $X$ respectively. 
For positive integer $n$,$i,j,k$, denote $[n]=\{1,2,\dots,n\}$, and $i\neq j\neq k$ means $i\neq j, j\neq k, k\neq i$. Given positive integer $t$, $\sum_{i_1\neq i_2\neq\dots\neq i_t}$ means summation over all integers $i_1,i_2,\dots,i_t$ in $[n]$ such that $|\{i_1,i_2,\dots,i_t\}|=t$. $\sum_{i_1< i_2<\dots< i_t}$ means summation over all integers $i_1,i_2,\dots,i_t$ in $[n]$ such that $i_1<i_2<\dots<i_t$. For a vector $W=(W_1,W_2,\dots,W_m)\in\mathbb{R}^m$ and a positive integer $q$, $||W||_q=\left(\sum_i^m|W_i|^q\right)^{\frac{1}{q}}$.

%%%%%%%%%%%%%%%%%%%%%%%%%%%%%%%%%%%%%%%%%%%%%%
%% Main Results                              
%%%%%%%%%%%%%%%%%%%%%%%%%%%%%%%%%%%%%%%%%%%%%%
\section{Main result}\label{mainRe}

In this paper, we study multilayer networks where all layers have the same set of nodes and edges only connect nodes within each layer \cite{D17,DP15,LLK20,CLM22,Arroyo21, PW24, Pensky24}. Specifically, the multilayer networks consists of  $L$ graphs $G_1,G_2,\dots,G_L$, with $G_l=(\mathcal{V},\mathcal{E}_l)$, where $\mathcal{V}=\{1,2,\dots,n\}$ is the node set, and $\mathcal{E}_l$ denotes a set of edges in graph $G_l$. Assume the graphs are undirected and unweighted, without self-loops. Each graph $G_l$ can be represented as a $n\times n$ symmetric adjacency matrix $A_l$, where $A_{l,ij}=1$ if $\{i,j\}\in \mathcal{E}_l$ and $A_{l,ij}=0$ otherwise. A typical example of the multilayer networks is brain networks, where nodes represent brain regions, and edges model interactions between two brain regions \cite{D17,DP15,SBL13}. Multilayer brain networks have the same nodes, and there is no edge connecting nodes of different networks. In this paper, we consider the following model for the multilayer networks.

\begin{Definition}[Random Multilayer 
 Heterogeneous Graphs]
 Given a positive integer $L$,
let $W_l$ be a vector in $[0,1]^n$ such that $||W_l||_2=1$ for all $l\in[L]$, and $\rho_{l}$ be a positive sequence that may depend on $n$. We say the multilayer networks $A_1,A_2,\dots, A_L$ follow the Random Multilayer 
 Heterogeneous Graphs Model $\mathcal{G}_{n}(W_1,W_2,\dots,W_L)$ if
 \[\mathbb{P}(A_{l,ij}=1)=\rho_{l}W_{l;i}W_{l;j},\ \ \ i<j,\]
where $A_{l,ii}=0$, $A_{l,ij}=A_{l,ji}$, $A_{l,ij}$ ($1\leq i<j\leq n$, $1\leq l\leq L$) are independent.
\end{Definition}

Each layer of network in $\mathcal{G}_{n}(W_1,W_2,\dots,W_L)$ is the degree-corrected  Erd\H{o}s-R\'{e}nyi random graph \cite{ZLZ12,GMZZ18}.
The random multilayer graphs $\mathcal{G}_{n}(W_1,W_2,\dots,W_L)$ is related to the models defined in \cite{Arroyo21, PW24, Pensky24}. If $W_1=W_2=\dots=W_L$, then $\mathcal{G}_{n}(W_1,W_2,\dots,W_L)$ is a special case of the common subspace independent edge random graph model in \cite{Arroyo21}. In this case, the multilayer networks $A_l$ ($1\leq l\leq L$) share the same subspace represented by $W$, and simultaneously  have sufficient heterogeneity due to distinct $\rho_l$. \cite{Arroyo21} proposed consistent estimators of the common subspace and heterogeneity parameters $\rho_l$.
When some of the vectors $W_1,W_2,\dots, W_L$ are equal, $\mathcal{G}_{n}(W_1,W_2,\dots,W_L)$ is a special case of the diverse multilayerx networks model in \cite{PW24, Pensky24}. In this scenario, the multilayer networks $A_l$ ($1\leq l\leq L$) can be partitioned into clusters such that the networks within the same cluster have common subspaces. \cite{PW24, Pensky24} presented several algorithms to recover the latent cluster and estimate the common subspaces.

The estimation methods in \cite{Arroyo21,PW24, Pensky24}  depend on the assumption that some or all of $W_l$ are equal. In practice, it is unknown whether some or all of the multilayer networks share common subspaces represented by $W_l$. In this paper, we first attempt to solve this problem by means of hypothesis testing.
Given multilayer networks $A_1,A_2,\dots, A_L\sim\mathcal{G}_{n}(W_1,W_2,\dots,W_L)$, we are interesting in testing the following hypotheses
\begin{equation}\label{hypoeq}
H_0:W_1=W_2=\dots=W_L,\ \ \hskip 1cm H_1: W_{l_1}\neq W_{l_2},\ \text{for some }  l_1\neq l_2.
\end{equation}
Under $H_0$, the expected adjacency matrices of the graphs $A_1,A_2,\dots, A_L$ have a  common subspace. This implies that the graphs can be embedded into the same latent subspace. Under $H_1$, there exist at least two graphs such that their embedding subspaces are  different.

We propose the first test, called the Weighted Degree Difference Test, for the hypothesis testing problem (\ref{hypoeq}). For each $l\in[L]$,
let $P_{l}=\sum_{i\neq j\neq k}A_{l,ij}A_{l,jk}$, $d_{l;i}=\sum_{j}A_{l,ij}$ and $d_l=\sum_{i,j}A_{l,ij}$. Define the Weighted Degree Difference Test statistic $D_n$ as
\begin{equation}\label{testst0}
D_n=\frac{1}{\sigma_n}\sum_{l=2}^{L}\left[\sum_{i=1}^n\left(\frac{d_{1,i}}{\sqrt{P_1}}-\frac{d_{l,i}}{\sqrt{P_l}}\right)^2-\frac{d_{1}}{P_1}-\frac{d_{l}}{P_l}\right],
\end{equation}
where 
\begin{equation}\label{varsigma}
\sigma_n^2=\frac{2(L-1)^2}{P_1}+\sum_{l=2}^L\frac{2}{P_l}+\sum_{l=2}^L\frac{4}{\sqrt{P_1P_l}}.
\end{equation}
In (\ref{testst0}), the term $\frac{d_{1,i}}{\sqrt{P_1}}-\frac{d_{l,i}}{\sqrt{P_l}}$ is a weighted difference between degree $d_{1,i}$ of node $i$ in $A_1$ and $d_{l,i}$ of node $i$ in $A_l$. Hence we call $D_n$ the Weighted Degree Difference Test statistic.  Under $H_0$, the limiting distribution of $D_n$ is the standard normal distribution as shown below.

\begin{Theorem}\label{mainthm0} 
Let $L$ be a fixed positive integer.
Suppose $\min_{1\leq l\leq L}\{\rho_l\}=\omega(1)$,    $W_l=W$ for all $l\in\{1,2,\dots,L\}$,
 $||W||_1=\omega(1)$,  $\rho_l=o(||W||_1)$ and $||W||_4=o(1)$.
    Under $H_0$, $D_n$ converges in distribution to the standard normal distribution, as $n$ goes to infinity.
\end{Theorem}

Based on Theorem \ref{mainthm0}, we define the Weighted Degree Difference Test (WDDT) as follows:
\[\text{Reject $H_0$ at significance level $\alpha$, if $|D_n|>Z_{\frac{\alpha}{2}}$},\]
where $Z_{\frac{\alpha}{2}}$ is the $100(1-\frac{\alpha}{2})\%$ quantile of the standard normal distribution. Theorem \ref{mainthm0} guarantees the type I error of the WDDT test is asymptotically equal to $\alpha$.

The expected total degree of network $A_l$ is equal to $\rho_l||W||_1^2$.
The conditions $\min_{1\leq l\leq L}\{\rho_l\}=\omega(1)$  in Theorem \ref{mainthm0} and  $||W||_1=\omega(1)$ indicate each layer of network should have growing total degree. Similar assumption is needed in testing community structure in networks \cite{GMZZ18,JKL21,L16}. 
The conditions $\rho_l=o(||W||_1)$ and $||W||_4=o(1)$ require that the networks should not be dense. Since many real networks are sparse, this assumption is not restrictive \cite{A17}.

\begin{Theorem}\label{mainthm1} 
Let $L$ be a fixed positive integer.
Assume $\min_{1\leq l\leq L}\{\rho_l\}=\omega(1)$,  $\rho_l=o(||W_l||_1)$,
 and $||W_{l}||_4=o(1)$ for all $l\in\{1,2,\dots,L\}$.
    Suppose  there exists $l_0\in\{2,\dots,L\}$ such that $\sum_iW_{1;i}W_{l_0;i}\leq 1-\epsilon$ for some constant $\epsilon\in(0,1)$. Under $H_1$, the WDDT test statistics $D_n$ has the following lower bound
    \[D_n\geq \Theta\left(\frac{1}{r_n}\sum_{l=2}^L\left(1-\sum_{i=1}^nW_{1,i}W_{l,i}\right)\right)(1+o_P(1)),\]
    where
    \begin{eqnarray} \label{sigmaMorder}
r_n^2&=&\frac{2(L-1)^2}{\rho_1^2||W_1||_1^2}+\sum_{l=2}^L\frac{2}{\rho_l^2||W_l||_1^2}+\sum_{l=2}^L\frac{4}{\rho_1||W_1||_1\rho_l||W_l||_1}.
\end{eqnarray}

\end{Theorem}

Note that 
\begin{equation}\label{cheq1}
\sum_iW_{1;i}W_{l;i}\leq \sqrt{\sum_iW_{1;i}^2}\sqrt{\sum_iW_{l;i}^2}=1,
\end{equation}
where equality holds if and only if $W_1=c W_l$ for some constant $c$. Since $||W_l||_2=1$ and $W_l\in[0,1]^n$, then $c=1$. Hence the equality in (\ref{cheq1}) holds if and only if $W_1=W_l$. Under $H_0$, $W_1=W_l$ for all $l\in[L]$. Then $D_n=O_P(1)$ by Theorem \ref{mainthm0}. Under $H_1$, there exists $l_0\neq 1$ such that $W_{l_0}\neq W_1$. If $\sum_iW_{1;i}W_{l_0;i}\leq 1-\epsilon$ for a constant $\epsilon\in(0,1)$, then
    \[D_n\geq \Theta\left(\frac{\epsilon}{r_n}\right)(1+o_P(1)).\]
Under the assumption of Theorem \ref{mainthm1}, $D_n$ goes to infinity in probability.  In this case, the power of the WDDT goes to one as $n$ goes to infinity.

The assumptions of Theorem \ref{mainthm0} and Theorem \ref{mainthm1} are not restrictive. We provide two examples below.

\begin{Example}\label{example0}
For positive constants $r,\lambda_l$ with $\lambda_l\leq1$ and $r>1$,
    let $W_{l,i}=\frac{\lambda_l\sqrt{r}}{\sqrt{n}}$ for $1\leq i\leq \frac{n}{r}$ and $W_{l,i}=\frac{\sqrt{\frac{r}{r-1}(1-\lambda_l^2)}}{\sqrt{n}}$ for $\frac{n}{r}<i\leq n$. Then simple calculation yields 
    \[||W_l||_1=\Theta(\sqrt{n}),\ \ \ ||W_l||_2=1,\ \ \ \ ||W_l||_4^4=O\left(\frac{1}{n}\right).\]        
    If $\lambda_l=\lambda_1$, $\min_{1\leq l\leq L}\{\rho_l\}=\omega(1)$,  $\rho_l=o(\sqrt{n})$, all the conditions of Theorem \ref{mainthm0}  are satisfied. 

 Moreover, direct calculation yields
 \begin{eqnarray*}
\sum_{i=1}^nW_{1,i}W_{l,i}
&=&\lambda_1\lambda_l+\sqrt{(1-\lambda_1^2)(1-\lambda_l^2)}.
\end{eqnarray*} 
If $\lambda_l\neq\lambda_1$,  there exists a positive constant $\epsilon$ such that $\sum_{i=1}^nW_{1,i}W_{l,i}\leq 1-\epsilon$. The larger the difference between $\lambda_1$ and $\lambda_l$, the smaller the $\sum_{i=1}^nW_{1,i}W_{l,i}$.  Hence the assumptions of Theorem \ref{mainthm1} are satisfied.

\end{Example}

\begin{Example}\label{example1}
    Let $m,\beta_l$ be non-negative constants for $l\in[L]$. Denote
    \[S_{n,m}=\sum_{i=1}^ni^{m}.\]
    When $m$ is a positive integer, $S_{n,m}$ is given by the Faulhaber's formula. For arbitrary positive constant $m$,
it is easy to verify that 
    \begin{equation}\label{sumnm}
        S_{n,m}=\frac{n^{m+1}}{m+1}\left(1+o(1)\right).
    \end{equation}

    Let 
    \[W_{l,i}=\frac{i^{\beta_l}}{\sqrt{S_{n,2\beta_l}}}.\]    
    By the definition of $S_{n,m}$ and (\ref{sumnm}), one has
    \[||W_l||_1=\sum_{i=1}^nW_{l,i}=\frac{1}{\sqrt{S_{n,2\beta_l}}}\sum_{i=1}^ni^{\beta_l}=\Theta\left(\sqrt{n}\right),\]
    \[||W_l||_2^2=\sum_{i=1}^nW_{l,i}^2=\frac{1}{S_{n,2\beta_l}}\sum_{i=1}^ni^{2\beta_l}=1,\]        \[||W_l||_4^4=\sum_{i=1}^nW_{l,i}^4=\frac{1}{S_{n,2\beta_l}^2}\sum_{i=1}^ni^{4\beta_l}=O\left(\frac{1}{n}\right).\]
    The expected degree of node $i$ in $A_l$ is $i^{\beta_l}\Theta\left(\frac{\rho_l}{n^{\beta_l}}\right)$. Hence the networks are highly heterogeneous.
    If $\beta_l=\beta_1$, $\min_{1\leq l\leq L}\{\rho_l\}=\omega(1)$,   $\rho_l=o(\sqrt{n})$, all the conditions of Theorem \ref{mainthm0}  are satisfied. 

In addition, direct calculation yields
\begin{eqnarray*}
\sum_{i=1}^nW_{1,i}W_{l,i}&=&\sum_{i=1}^n\frac{i^{\beta_1}}{\sqrt{S_{n,2\beta_1}}}\frac{i^{\beta_l}}{\sqrt{S_{n,2\beta_l}}}\\
&=&(1+o(1))\frac{n^{\beta_1+\beta_l+1}}{\beta_1+\beta_l+1}\frac{\sqrt{(2\beta_1+1)(2\beta_l+1)}}{\sqrt{n^{2\beta_1+2\beta_l+2}}}\\
&=&(1+o(1))\frac{\sqrt{(2\beta_1+1)(2\beta_l+1)}}{\beta_1+\beta_l+1}.
\end{eqnarray*}
    If $\beta_l\neq \beta_1$, there exists a positive constant $\epsilon$ such that $\sum_{i=1}^nW_{1,i}W_{l,i}\leq 1-\epsilon$. The larger the difference between $\beta_1$ and $\beta_l$, the smaller the $\sum_{i=1}^nW_{1,i}W_{l,i}$.  Hence the assumptions of Theorem \ref{mainthm1} are satisfied.
\end{Example}

%%%%%%%%%%%%%%%%%%%%%%%%%%%%%%%%%%%%%%%%%%%%%%
%% Simulation and real data application
%%%%%%%%%%%%%%%%%%%%%%%%%%%%%%%%%%%%%%%%%%%%%%

\section{Simulation and real data application}\label{SimRD}

In this section, we study the performance of the proposed WDDT test on simulated networks and apply the test to a real-world multilayer networks.

\subsection{Simulation}

In this simulation study, the nominal type I error is 0.05. We repeat the simulation 1000 times to calculate empirical type I errors and powers. We take the number of networks $L\in\{2, 3, 4\}$, the network size $n\in\{200, 250, 300\}$, and $\rho_l=n^{\tau_l}$ with $\boldsymbol{\tau}=(\tau_1,\tau_2,\tau_3,\tau_4)=(0.3,0.2,0.4,0.1)$. 

Firstly, we consider simulation setting  specified in Example \ref{example0}. That is, the vectors $W_l=(W_{l,1},W_{l,2},\dots,W_{l,n})$ ($1\leq l\leq L$) are given by 
 \[W_{l,i}=\frac{\lambda_l\sqrt{r}}{\sqrt{n}},\ \ \ 1\leq i\leq \frac{n}{r},\]
\[W_{l,i}=\frac{\sqrt{\frac{r}{r-1}(1-\lambda_l^2)}}{\sqrt{n}},\ \ \ \frac{n}{r}<i\leq n,\]
where $0<\lambda_l\leq 1$ and $r>1$. Then network $A_l$ is generated as follows 
 \begin{equation}\label{Simeq1}
 \mathbb{P}(A_{l,ij}=1)=\rho_{l}W_{l;i}W_{l;j},\ \ \ i<j,
 \end{equation}
where $A_{l,ij}=A_{l,ji}$,  $A_{l,ij}$ ($1\leq i<j\leq n$, $1\leq l\leq L$) are independent. We take $r\in\{2,2.5,3\}$. Denote $\boldsymbol{\tau}=(\tau_1,\dots,\tau_L)$ and $\boldsymbol{\lambda}=(\lambda_1,\dots,\lambda_L)$ with $\lambda_l\in\{0.8,0.7,0.6,0.5\}$.

Under $H_0$, $\lambda_1=\dots=\lambda_L=0.8$. That is, all the $\lambda_l$ are equal, and hence $W_l$ are the same. Under $H_1$, we consider two cases. The first case corresponds to $\lambda_1=0.8$ and $\lambda_2=\dots=\lambda_L\in\{0.7,0.6,0.5\}$. The simulation results are summarized in Table \ref{exsimu1}. In the second case, $\lambda_1=0.8$, $\lambda_2,\dots,\lambda_L\in\{0.7,0.6,0.5\}$ and $\lambda_2,\dots,\lambda_L$ are not the same. The simulation results are presented in Table \ref{exsimu2}.

The simulated type I errors are listed in the second column of Table \ref{exsimu1} and Table \ref{exsimu2}. All the type I errors are close to 0.05. This result indicates Theorem \ref{mainthm0} works for small netowrk size $n$. For fixed $r, L,\boldsymbol{\lambda}$, the power increases as the network size $n$ increases. For fixed $r, L,n$, the power increases as the difference between  $\boldsymbol{\lambda}$  and $\boldsymbol{\lambda}_0=(0.8,\dots,0.8)$ gets larger. Moreover, the maximum power is almost one. These findings are consistent with Theorem \ref{mainthm1}.

In the second simulation, the networks are generated from the model specified in Example \ref{example1}. That is, the vectors $W_l=(W_{l,1},W_{l,2},\dots,W_{l,n})$ ($1\leq l\leq L$) are given by 
\[W_{l,i}=\frac{i^{\beta_l}}{\sqrt{S_{n,2\beta_l}}},\] 
where $S_{n,m}=\sum_{i=1}^ni^{m}$, $m$ and $\beta_l$ are non-negative constants for $l\in[L]$. Let $\boldsymbol{\beta}=(\beta_1,\beta_2,\beta_3,\beta_4)$ with $\beta_l\in\{1,2,3,4\}$.
Then generate $A_l$ according to (\ref{Simeq1}). 

Under $H_0$,  $\beta_1=\dots=\beta_L=1$. In this case, $W_1=\dots=W_L$. Under $H_1$, there are two scenarios. In the first scenario, $\beta_1=1$ and $\beta_2=\dots=\beta_L\in\{2,3,4\}$. Table \ref{betata1} reports the results. In the second scenario, $\beta_1=1$, $\beta_2,\dots,\beta_L\in\{2,3,4\}$ and  $\beta_2,\dots,\beta_L$ are not equal. Table \ref{betata2} reports the results.

The second columns in Table \ref{betata1} and Table \ref{betata2} show all the type I errors are close to 0.05. This result indicates Theorem \ref{mainthm0} works for small netowrk size $n$. For fixed $ L,\boldsymbol{\beta}$, the power increases as the network size $n$ increases. For fixed $L,n$, the power increases as the difference between  $\boldsymbol{\beta}$  and $\boldsymbol{\beta}_0=(1,\dots,1)$ gets larger. Moreover, the maximum power is one. These findings are consistent with Theorem \ref{mainthm1}.

In summary, this simulation study shows the proposed WDDT test has satisfactory performance.

\begin{table}[]

\centering
%\begin{adjustbox}{width=1\textwidth}
\small

\begin{tabular}{|c|| c| c| c |c|}
 \hline
 \multicolumn{5}{|c|}{$r = 2$} \\
 \hline
 \multicolumn{5}{|c|}{$L=2, \boldsymbol{\tau}=({0.3, 0.2})$} \\
 \hline
 $  $ & $\boldsymbol{\lambda} = ({0.8, 0.8})$ &$\boldsymbol{\lambda} = ({0.8, 0.7})$&$\boldsymbol{\lambda} = ({0.8, 0.6})$& $\boldsymbol{\lambda} =({0.8, 0.5})$\\
 \hline
$n=200$&$0.040$ &$0.040$ &$0.221$ &$0.671$ \\
\hline
$n=250$&$0.038$ &$0.062$ &$0.325$ &$0.815$ \\
\hline
$n=300$&$0.049$ &$0.066$ &$0.407$ &$0.913$ \\
\hline

\multicolumn{5}{|c|}{$L=3, \boldsymbol{\tau}=({0.3, 0.2, 0.4})$} \\
 \hline
 $  $ &$\boldsymbol{\lambda} = ({0.8, 0.8, 0.8})$ &$\boldsymbol{\lambda} = ({0.8, 0.7, 0.7})$&$\boldsymbol{\lambda} =({0.8, 0.6, 0.6})$& $\boldsymbol{\lambda} =({0.8, 0.5, 0.5})$\\
 \hline
$n=200$&$0.052$ &$0.056$ &$0.467$ &$0.957$ \\
\hline
$n=250$&$0.053$ &$0.077$ &$0.592$ &$0.984$ \\
\hline
$n=300$&$0.054$ &$0.088$ &$0.750$ &$0.999$ \\
\hline

\multicolumn{5}{|c|}{$L=4, \boldsymbol{\tau}=({0.3, 0.2, 0.4, 0.1})$} \\
 \hline
 $  $ &$\boldsymbol{\lambda} = ({0.8, 0.8, 0.8, 0.8})$ &$\boldsymbol{\lambda} = ({0.8, 0.7, 0.7, 0.7})$&$\boldsymbol{\lambda} = ({0.8, 0.6, 0.6, 0.6})$& $\boldsymbol{\lambda} = ({0.8, 0.5, 0.5, 0.5})$\\
 \hline
$n=200$&$0.059$ &$0.062$ &$0.385$ &$0.918$ \\
\hline
$n=250$&$0.053$ &$0.066$ &$0.503$ &$0.981$ \\
\hline
$n=300$&$0.048$ &$0.084$ &$0.612$ &$0.995$ \\
\hline
\end{tabular}
%\end{adjustbox}
%%%%%%%%%%%%%%%%%%%%%%%%%%%%%%%%%%%%%%%%%%%%%%%%%%%%%%%%%%%

%\hline
\centering
%\begin{adjustbox}{width=1\textwidth}
\small
\begin{tabular}{|c|| c| c| c |c|}
 \hline
 \multicolumn{5}{|c|}{$r = 2.5$} \\
 \hline
\multicolumn{5}{|c|}{$L=2, \boldsymbol{\tau}= ({0.3, 0.2})$} \\
 \hline
 $  $ &$\boldsymbol{\lambda} = ({0.8, 0.8})$ &$\boldsymbol{\lambda} = ({0.8, 0.7})$&$\boldsymbol{\lambda} = ({0.8, 0.6})$& $\boldsymbol{\lambda} =({0.8, 0.5})$\\
 \hline
$n=200$&$0.058$ &$0.046$ &$0.195$ &$0.663$ \\
\hline
$n=250$&$0.059$ &$0.055$ &$0.268$ &$0.823$ \\
\hline
$n=300$&$0.041$ &$0.077$ &$0.381$ &$0.899$ \\
\hline

\multicolumn{5}{|c|}{$L=3, \boldsymbol{\tau}=({0.3, 0.2, 0.4})$} \\
 \hline
 $  $ &$\boldsymbol{\lambda} = ({0.8, 0.8, 0.8})$ &$\boldsymbol{\lambda} = ({0.8, 0.7, 0.7})$&$\boldsymbol{\lambda} =({0.8, 0.6, 0.6})$& $\boldsymbol{\lambda} =({0.8, 0.5, 0.5})$\\
 \hline
$n=200$&$0.051$ &$0.045$ &$0.499$ &$0.965$ \\
\hline
$n=250$&$0.050$ &$0.065$ &$0.586$ &$0.991$ \\
\hline
$n=300$&$0.048$ &$0.086$ &$0.719$ &$0.999$ \\
\hline

\multicolumn{5}{|c|}{$L=4, \boldsymbol{\tau}=({0.3, 0.2, 0.4, 0.1})$} \\
 \hline
 $  $ &$\boldsymbol{\lambda} = ({0.8, 0.8, 0.8, 0.8})$ &$\boldsymbol{\lambda} = ({0.8, 0.7, 0.7, 0.7})$&$\boldsymbol{\lambda}= ({0.8, 0.6, 0.6, 0.6})$& $\boldsymbol{\lambda} = ({0.8, 0.5, 0.5, 0.5})$\\
 \hline
$n=200$&$0.056$ &$0.063$ &$0.372$ &$0.916$ \\
\hline
$n=250$&$0.056$ &$0.066$ &$0.502$ &$0.975$ \\
\hline
$n=300$&$0.058$ &$0.076$ &$0.651$ &$0.995$ \\
\hline
\end{tabular}
%\end{adjustbox}
%%%%%%%%%%%%%%%%%%%%%%%%%%%%%%%%%%%%%%%%%%%%%%%%%%%%%%%%%%%

%\hline
\centering
%\begin{adjustbox}{width=1\textwidth}
\small
\begin{tabular}{|c|| c| c| c |c|}
 \hline
 \multicolumn{5}{|c|}{$r = 3$} \\
 \hline
\multicolumn{5}{|c|}{$L=2, \boldsymbol{\tau}=({0.3, 0.2})$} \\
 \hline
 $  $ &$\boldsymbol{\lambda} = ({0.8, 0.8})$ &$\boldsymbol{\lambda} = ({0.8, 0.7})$&$\boldsymbol{\lambda} = ({0.8, 0.6})$& $\boldsymbol{\lambda} =({0.8, 0.5})$\\
 \hline
$n=200$&$0.038$ &$0.047$ &$0.176$ &$0.656$ \\
\hline
$n=250$&$0.054$ &$0.050$ &$0.247$ &$0.819$ \\
\hline
$n=300$&$0.043$ &$0.067$ &$0.344$ &$0.912$ \\
\hline

\multicolumn{5}{|c|}{$L=3, \boldsymbol{\tau}=({0.3, 0.2, 0.4})$} \\
 \hline
 $  $ &$\boldsymbol{\lambda} = ({0.8, 0.8, 0.8})$ &$\boldsymbol{\lambda} = ({0.8, 0.7, 0.7})$&$\boldsymbol{\lambda} =({0.8, 0.6, 0.6})$& $\boldsymbol{\lambda} =({0.8, 0.5, 0.5})$\\
 \hline
$n=200$&$0.056$ &$0.055$ &$0.421$ &$0.949$ \\
\hline
$n=250$&$0.055$ &$0.061$ &$0.589$ &$0.997$ \\
\hline
$n=300$&$0.054$ &$0.070$ &$0.705$ &$0.998$ \\
\hline

\multicolumn{5}{|c|}{$L=4, \boldsymbol{\tau}=({0.3, 0.2, 0.4, 0.1})$} \\
 \hline
 $  $ &$\boldsymbol{\lambda} = ({0.8, 0.8, 0.8, 0.8})$ &$\boldsymbol{\lambda} = ({0.8, 0.7, 0.7, 0.7})$&$\boldsymbol{\lambda} = ({0.8, 0.6, 0.6, 0.6})$& $\boldsymbol{\lambda} = ({0.8, 0.5, 0.5, 0.5})$\\
 \hline
$n=200$&$0.058$ &$0.061$ &$0.340$ &$0.928$ \\
\hline
$n=250$&$0.061$ &$0.066$ &$0.467$ &$0.981$ \\
\hline
$n=300$&$0.056$ &$0.073$ &$0.600$ &$0.998$ \\
\hline
\end{tabular}
%\end{adjustbox}
%%%%%%%%%%%%%%%%%%%%%%%%%%%%%%%%%%%%%%%%%%%%%%%%%%%%%%%%%%%
\caption{Simulation results based on Example 2.1: One Difference in $H_1$}\label{exsimu1}
\end{table}

\begin{table} 

\centering
%\begin{adjustbox}{width=1\textwidth}
\small

\begin{tabular}{ |c|| c| c| c |c|}
 \hline
 \multicolumn{5}{|c|}{$r = 2$} \\
 \hline
 \multicolumn{5}{|c|}{$L=2, \boldsymbol{\tau}=({0.3, 0.2})$} \\
 \hline
 $  $ &$\boldsymbol{\lambda} = ({0.8, 0.8})$ &$\boldsymbol{\lambda} = ({0.8, 0.7})$&$\boldsymbol{\lambda} = ({0.8, 0.6})$& $\boldsymbol{\lambda} =({0.8, 0.5})$\\
 \hline
$n=200$&$0.040$ &$0.040$ &$0.221$ &$0.671$ \\
\hline
$n=250$&$0.038$ &$0.062$ &$0.325$ &$0.815$ \\
\hline
$n=300$&$0.049$ &$0.066$ &$0.407$ &$0.913$ \\
\hline

\multicolumn{5}{|c|}{$L=3, \boldsymbol{\tau}=({0.3, 0.2, 0.4})$} \\
 \hline
 $  $ &$\boldsymbol{\lambda} = ({0.8, 0.8, 0.8})$ &$\boldsymbol{\lambda} = ({0.8, 0.7, 0.6})$&$\boldsymbol{\lambda} =({0.8, 0.7, 0.5})$& $\boldsymbol{\lambda} =({0.8, 0.6, 0.5})$\\
 \hline
$n=200$&$0.052$ &$0.186$ &$0.567$ &$0.785$ \\
\hline
$n=250$&$0.053$ &$0.246$ &$0.721$ &$0.899$ \\
\hline
$n=300$&$0.054$ &$0.368$ &$0.862$ &$0.972$ \\
\hline

\multicolumn{5}{|c|}{$L=4, \boldsymbol{\tau}=({0.3, 0.2, 0.4, 0.1})$} \\
 \hline
 $  $ &$\boldsymbol{\lambda} = ({0.8, 0.8, 0.8, 0.8})$ &$\boldsymbol{\lambda} = ({0.8, 0.7, 0.6, 0.6})$&$\boldsymbol{\lambda} = ({0.8, 0.7, 0.6, 0.5})$& $\boldsymbol{\lambda} = ({0.8, 0.7, 0.5, 0.5})$\\
 \hline
$n=200$&$0.059$ &$0.215$ &$0.499$ &$0.656$ \\
\hline
$n=250$&$0.053$ &$0.308$ &$0.604$ &$0.827$ \\
\hline
$n=300$&$0.048$ &$0.428$ &$0.721$ &$0.904$ \\
\hline
\end{tabular}
%\end{adjustbox}
%%%%%%%%%%%%%%%%%%%%%%%%%%%%%%%%%%%%%%%%%%%%%%%%%%%%%%%%%%%

%\hline
\centering
%\begin{adjustbox}{width=1\textwidth}
\small
\begin{tabular}{ |c|| c| c| c |c|}
 \hline
 \multicolumn{5}{|c|}{$r = 2.5$} \\
 \hline
\multicolumn{5}{|c|}{$L=2, \boldsymbol{\tau}=({0.3, 0.2})$} \\
 \hline
 $  $ &$\boldsymbol{\lambda} = ({0.8, 0.8})$ &$\boldsymbol{\lambda} = ({0.8, 0.7})$&$\boldsymbol{\lambda} = ({0.8, 0.6})$& $\boldsymbol{\lambda} =({0.8, 0.5})$\\
 \hline
$n=200$&$0.058$ &$0.046$ &$0.195$ &$0.663$ \\
\hline
$n=250$&$0.059$ &$0.055$ &$0.268$ &$0.823$ \\
\hline
$n=300$&$0.041$ &$0.077$ &$0.381$ &$0.899$ \\
\hline

\multicolumn{5}{|c|}{$L=3, \boldsymbol{\tau}=({0.3, 0.2, 0.4})$} \\
 \hline
 $  $ &$\boldsymbol{\lambda} = ({0.8, 0.8, 0.8})$ &$\boldsymbol{\lambda} = ({0.8, 0.7, 0.6})$&$\boldsymbol{\lambda} =({0.8, 0.7, 0.5})$& $\boldsymbol{\lambda} =({0.8, 0.6, 0.5})$\\
 \hline
$n=200$&$0.051$ &$0.167$ &$0.617$ &$0.796$ \\
\hline
$n=250$&$0.050$ &$0.266$ &$0.774$ &$0.925$ \\
\hline
$n=300$&$0.048$ &$0.353$ &$0.907$ &$0.969$ \\
\hline

\multicolumn{5}{|c|}{$L=4, \boldsymbol{\tau}=({0.3, 0.2, 0.4, 0.1})$} \\
 \hline
 $  $ &$\boldsymbol{\lambda} = ({0.8, 0.8, 0.8, 0.8})$ &$\boldsymbol{\lambda} = ({0.8, 0.7, 0.6, 0.6})$&$\boldsymbol{\lambda} = ({0.8, 0.7, 0.6, 0.5})$& $\boldsymbol{\lambda} = ({0.8, 0.7, 0.5, 0.5})$\\
 \hline
$n=200$&$0.056$ &$0.227$ &$0.447$ &$0.659$ \\
\hline
$n=250$&$0.056$ &$0.307$ &$0.607$ &$0.835$ \\
\hline
$n=300$&$0.058$ &$0.409$ &$0.719$ &$0.913$ \\
\hline
\end{tabular}
%\end{adjustbox}
%%%%%%%%%%%%%%%%%%%%%%%%%%%%%%%%%%%%%%%%%%%%%%%%%%%%%%%%%%%

%\hline
\centering
%\begin{adjustbox}{width=1\textwidth}
\small
\begin{tabular}{ |c|| c| c| c |c|}
 \hline
 \multicolumn{5}{|c|}{$r = 3$} \\
 \hline
\multicolumn{5}{|c|}{$L=2, \boldsymbol{\tau}=({0.3, 0.2})$} \\
 \hline
 $  $ &$\boldsymbol{\lambda} = ({0.8, 0.8})$ &$\boldsymbol{\lambda} = ({0.8, 0.7})$&$\boldsymbol{\lambda} = ({0.8, 0.6})$& $\boldsymbol{\lambda} =({0.8, 0.5})$\\
 \hline
$n=200$&$0.038$ &$0.047$ &$0.176$ &$0.656$ \\
\hline
$n=250$&$0.054$ &$0.050$ &$0.247$ &$0.819$ \\
\hline
$n=300$&$0.043$ &$0.067$ &$0.344$ &$0.912$ \\
\hline

\multicolumn{5}{|c|}{$L=3, \boldsymbol{\tau}=({0.3, 0.2, 0.4})$} \\
 \hline
 $  $ &$\boldsymbol{\lambda} = ({0.8, 0.8, 0.8})$ &$\boldsymbol{\lambda} = ({0.8, 0.7, 0.6})$&$\boldsymbol{\lambda} =({0.8, 0.7, 0.5})$& $\boldsymbol{\lambda} =({0.8, 0.6, 0.5})$\\
 \hline
$n=200$&$0.059$ &$0.167$ &$0.593$ &$0.811$ \\
\hline
$n=250$&$0.044$ &$0.252$ &$0.792$ &$0.935$ \\
\hline
$n=300$&$0.057$ &$0.378$ &$0.893$ &$0.985$ \\
\hline

\multicolumn{5}{|c|}{$L=4, \boldsymbol{\tau}=({0.3, 0.2, 0.4, 0.1})$} \\
 \hline
 $  $ &$\boldsymbol{\lambda} = ({0.8, 0.8, 0.8, 0.8})$ &$\boldsymbol{\lambda} = ({0.8, 0.7, 0.6, 0.6})$&$\boldsymbol{\lambda} = ({0.8, 0.7, 0.6, 0.5})$& $\boldsymbol{\lambda} = ({0.8, 0.7, 0.5, 0.5})$\\
 \hline
$n=200$&$0.058$ &$0.210$ &$0.424$ &$0.687$ \\
\hline
$n=250$&$0.056$ &$0.280$ &$0.594$ &$0.846$ \\
\hline
$n=300$&$0.055$ &$0.367$ &$0.719$ &$0.923$ \\
\hline
\end{tabular}
%\end{adjustbox}
%%%%%%%%%%%%%%%%%%%%%%%%%%%%%%%%%%%%%%%%%%%%%%%%%%%%%%%%%%%
\caption{Simulation results based on Example 2.1: More Than One Difference in $H_1$} \label{exsimu2}
\end{table}

%%%%%%%%%%%%%%%%%%%%%%%%%%%%%%%%%%%%%%%%%%%%%%
%% Table         
%%%%%%%%%%%%%%%%%%%%%%%%%%%%%%%%%%%%%%%%%%%%%%
\newpage

\begin{table}

\centering
%\begin{adjustbox}{width=1\textwidth}
%\normalsize

\begin{tabular}{ |c|| c| c| c |c|}
 \hline
 \multicolumn{5}{|c|}{$L=2, \boldsymbol{\tau}=({0.3, 0.2})$} \\
 \hline
 $  $ &$\boldsymbol{\beta} = ({1, 1})$ &$\boldsymbol{\beta} =({1, 2})$ &$\boldsymbol{\beta} = ({1, 3})$& $\boldsymbol{\beta} = ({1, 4})$\\
 \hline
$n=200$&$0.047$ &$0.096$ &$0.439$ &$0.756$ \\
\hline
$n=250$&$0.046$ &$0.119$ &$0.547$ &$0.906$ \\
\hline
$n=300$&$0.042$ &$0.149$ &$0.720$ &$0.962$ \\
\hline

\multicolumn{5}{|c|}{$L=3, \boldsymbol{\tau}=({0.3, 0.2, 0.4})$} \\
 \hline
 $  $ &$\boldsymbol{\beta} = ({1, 1, 1})$ &$\boldsymbol{\beta} = ({1, 2, 2})$&$\boldsymbol{\beta} = ({1, 3, 3})$& $\boldsymbol{\beta} = ({1, 4, 4})$\\
 \hline
$n=200$&$0.056$ &$0.157$ &$0.789$ &$0.990$ \\
\hline
$n=250$&$0.058$ &$0.210$ &$0.933$ &$0.998$ \\
\hline
$n=300$&$0.056$ &$0.320$ &$0.975$ &$1.000$ \\
\hline

\multicolumn{5}{|c|}{$L=4, \boldsymbol{\tau}=({0.3, 0.2, 0.4, 0.1})$} \\
 \hline
 $  $ &$\boldsymbol{\beta} = ({1, 1, 1, 1})$ &$\boldsymbol{\beta} = ({1, 2, 2, 2})$&$\boldsymbol{\beta} = ({1, 3, 3, 3})$& $\boldsymbol{\beta} = ({1, 4, 4, 4})$\\
 \hline
$n=200$&$0.054$ &$0.106$ &$0.709$ &$0.964$ \\
\hline
$n=250$&$0.060$ &$0.165$ &$0.850$ &$0.994$ \\
\hline
$n=300$&$0.057$ &$0.253$ &$0.933$ &$0.998$ \\
\hline
\end{tabular}
%\end{adjustbox}

\caption{Simulation results based on Example \ref{example1}: One Difference in $H_1$} \label{betata1}
\end{table}
%%%%%%%%%%%%%%%%%%%%%%%%%%%%%%%%%%%%%%%%%%%%%%%%%%%%%%%%%%%

%%%%%%%%%%%%%%%%%%%%%%%%%%%%%%%%%%%%%%%%%%%%%%
%% Table          
%%%%%%%%%%%%%%%%%%%%%%%%%%%%%%%%%%%%%%%%%%%%%%
\begin{table}
\centering
%\begin{adjustbox}{width=1\textwidth}
%\normalsize

\begin{tabular}{ |c|| c| c| c |c|}
 \hline
\multicolumn{5}{|c|}{$L=2, \boldsymbol{\tau}=({0.3, 0.2})$} \\
 \hline
 $  $ &$\boldsymbol{\beta} = ({1, 1})$ &$\boldsymbol{\beta} =({1, 2})$ &$\boldsymbol{\beta} = ({1, 3})$& $\boldsymbol{\beta} = ({1, 4})$\\
 \hline
$n=200$&$0.047$ &$0.096$ &$0.439$ &$0.756$ \\
\hline
$n=250$&$0.046$ &$0.119$ &$0.547$ &$0.906$ \\
\hline
$n=300$&$0.042$ &$0.149$ &$0.720$ &$0.962$ \\
\hline

\multicolumn{5}{|c|}{$L=3, \boldsymbol{\tau}=({0.3, 0.2, 0.4})$} \\
 \hline
 $  $ &$\boldsymbol{\beta} = ({1, 1, 1})$ &$\boldsymbol{\beta} = ({1, 2, 3})$&$\boldsymbol{\beta} = ({1, 2, 4})$& $\boldsymbol{\beta} = ({1, 3, 4})$\\
 \hline
$n=200$&$0.056$ &$0.501$ &$0.851$ &$0.953$ \\
\hline
$n=250$&$0.058$ &$0.717$ &$0.942$ &$0.994$ \\
\hline
$n=300$&$0.056$ &$0.807$ &$0.993$ &$0.998$ \\
\hline

\multicolumn{5}{|c|}{$L=4, \boldsymbol{\tau}=({0.3, 0.2, 0.4, 0.1})$} \\
 \hline
 $  $ &$\boldsymbol{\beta} = ({1, 1, 1, 1})$ &$\boldsymbol{\beta} = ({1, 2, 3, 3})$ & $\boldsymbol{\beta} = ({1, 2, 3, 4})$ &$\boldsymbol{\beta} = ({1, 2, 4, 4})$\\
 \hline
$n=200$&$0.054$ &$0.482$ &$0.672$ &$0.829$ \\
\hline
$n=250$&$0.060$ &$0.657$ &$0.846$ &$0.944$ \\
\hline
$n=300$&$0.057$ &$0.796$ &$0.918$ &$0.983$ \\
\hline
\end{tabular}
%\end{adjustbox}

\caption{Simulation results based on Example \ref{example1}: More Than One Difference in $H_a$} \label{betata2}
\end{table}

\subsection{Real data application}

In this subsection, we apply the proposed WDDT test to the multilayerx social networks CS-Aarhus  available in \cite{MMR13}. The CS-Aarhus networks are undirected and unweighted, and consist of five online and offline
relationships between the 61 employees
of Computer Science department at Aarhus. 
The first network $A_1$ represents two individuals having lunch together. Second network $A_2$ represents two individuals having a social connection via Facebook. The third network $A_3$ models co-authoring relationship among individuals. The fourth network $A_4$ represents two individuals having leisure together. The fifth network $A_5$ represents individuals working together.  The 5 networks $A_1, A_2, A_3, A_4, A_5$ and theire edge densities are listed in Table \ref{redata2} and visualized in Figure \ref{redata1}. Network $A_3$ is very sparse and networks $A_1$ and $A_5$ are relatively dense.

Firstly, we test whether the five networks share the same common invariant subspace. We calculate the WDDT test statistic $D_n$ and the corresponding $p$-value. The results are shown in Table \ref{redata3}.  The $p$-value is smaller than 0.05. This indicates the five networks do not share the same subspace.

Next, we test whether each quadripartite of the networks share the same common invariant subspace. By Table \ref{redata3}, the $p$-values are also smaller than 0.05. Every quadripartite of the networks cannot be embedded into the same subspace.

Further, we test whether each triple of the  networks share the same common invariant subspace. In this case, only the $p$-value for the triple $(A_3, A_4, A_5)$ exceeds 0.05. The networks $A_3, A_4, A_5$ can be embedded into the same subspace. Individuals co-authoring a publication ($A_3$) will probably have leisure together ($A_4$)  and work together ($A_5$).

Lastly, we test whether each pair of the networks have the same invariant subspace. As indicated in the above case, 
the $p$-values for testing the pairs $(A_3, A_4)$ and  $(A_3, A_5)$ are larger than 0.05. Moreover, the $p$-value for testing the pair $(A_1, A_4)$ exceeds 0.05. We fail to reject the null hypothesis that two networks have the same invariant subspace. 
The $p$-values for testing the remaining pairs $(A_1, A_2)$, $(A_1, A_3)$, $(A_1, A_5)$, $(A_2, A_3)$, $(A_2, A_4)$, $(A_2, A_5)$, $(A_4, A_5)$ are less than 0.05. Each pair does not share common invariant subspace. Note that in the previous test, $(A_3, A_4, A_5)$ shares the same invariant subspace. However, the $p$-value for $(A_4, A_5)$ is less than 0.05. This is a contradiction. The possible reason is that co-authoring network $A_3$ is extremely sparse (edge density 0.011). The test result involving $A_3$ may not be reliable.

\begin{table}[ht]
%\centering
%\begin{adjustbox}{width=1\textwidth}

\begin{tabular}{ |c|| l |c|  c|}
\hline
Layers& Association & Density \\
\hline
$A_1$ &  Representative of two individuals having Lunch together & 0.105  \\
\hline
$A_2$ &  Representative of two individuals having a social connection via Facebook & 0.068  \\
\hline
$A_3$ &  Representative of two individuals co-authoring a publication & 0.011  \\
\hline
$A_4$ &  Representative of two individuals having Leisure together & 0.048  \\
\hline
$A_5$ &  Representative of two individuals working together & 0.106  \\
\hline

\end{tabular}
%\end{adjustbox}

\caption{5 Layers of Real-World Network} \label{redata2}
\end{table}

%%%%%%%%%%%%%%%%%%%%%%%%%%%%%%%%%%%%%%%%%%%%%%
%% Table: Real Data PLOT  
%%%%%%%%%%%%%%%%%%%%%%%%%%%%%%%%%%%%%%%%%%%%%%
%%%%%%%%%%%%%%%%%%%%%%%%%%%%%%%%%%%%%%%%%%%%

%%%%%%%%%%%%%%%%%%%%%%%%%%%%%%%%%%%%%%%%%%%%%%
%% Table: Networks Graphs         
%%%%%%%%%%%%%%%%%%%%%%%%%%%%%%%%%%%%%%%%%%%%%%

\begin{table} 
\centering
%\begin{adjustbox}{width=1\textwidth}
\normalsize

\begin{tabular}{ |c| c|}
\hline
\includegraphics[width=7cm,height=6cm]{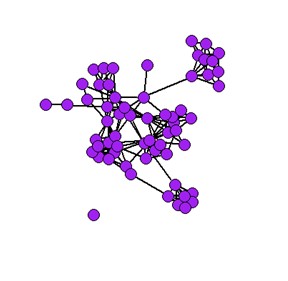} & \includegraphics[width=7cm,height=6cm]{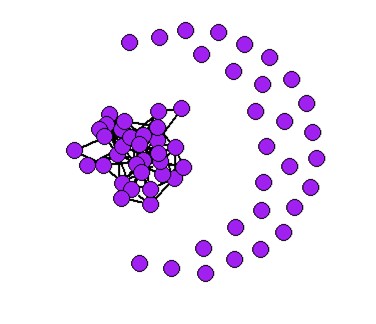} \\
\hline
$A_1$: Lunch & $A_2$: Facebook \\
\hline
\includegraphics[width=7cm,height=6cm]{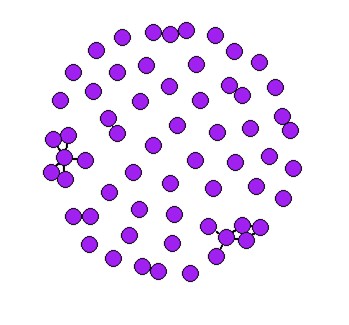} & \includegraphics[width=7cm,height=6cm]{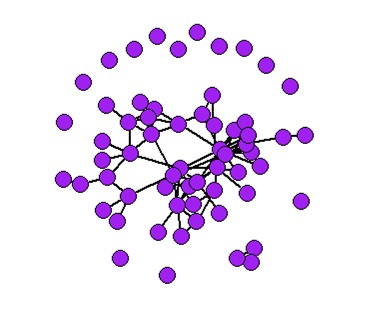} \\
\hline
$A_3$: Co-Authorship & $A_4$: Leisure \\
\hline
\includegraphics[width=7cm,height=6cm]{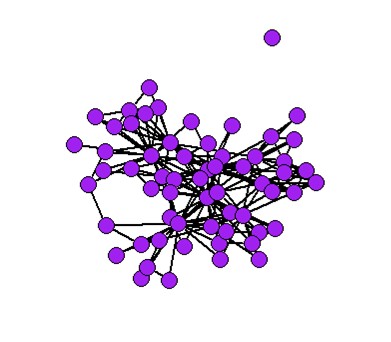} &  \multicolumn{1}{c}{} \\
\cline{1-1}
$A_5$: Work & \multicolumn{1}{c}{} \\
\cline{1-1}
\end{tabular}
%\end{adjustbox}

\caption{5 Layers of Real-World Network}  \label{redata1}
\end{table}
%%%%%%%%%%%%%%%%%%%%%%%%%%%%%%%%%%%%%%%%%%%%%%%%%%%%%%%%%%%

%%%%%%%%%%%%%%%%%%%%%%%%%%%%%%%%%%%%%%%%%%%%%%
%% Table: Real Data TABLE       
%%%%%%%%%%%%%%%%%%%%%%%%%%%%%%%%%%%%%%%%%%%%%%

\begin{table} 
\centering
%\begin{adjustbox}{width=1\textwidth}
%\large

\begin{tabular}{ |c|| c| c| c |c|}
\hline
Layers & Test Statistic & P-Value & Conclusion \\
\hline
%\hline
 $A_1, A_2$ & $3.641$ & $0.000$ & $Reject H_0$ \\
\hline
 $A_1, A_3$ & $2.472$ & $0.013$ & $Reject H_0$ \\
\hline
$A_1, A_4$ & $1.811$ & $0.070$ & $Not Reject H_0$ \\
\hline
 $A_1, A_5$ & $2.234$ & $0.025$ & $Reject H_0$ \\
\hline
$A_2, A_3$ & $2.344$ & $0.019$ & $Reject H_0$ \\
\hline
 $A_2, A_4$ & $3.816$ & $0.000$ & $Reject H_0$ \\
\hline
 $A_2, A_5$ & $3.233$ & $0.001$ & $Reject H_0$ \\
\hline
 $A_3, A_4$ & $0.818$ & $0.413$ & $Not Reject H_0$ \\
\hline
 $A_3, A_5$ & $0.819$ & $0.413$ & $Not Reject H_0$ \\
\hline
 $A_4, A_5$ & $3.723$ & $0.000$ & $Reject H_0$ \\
\hline
\hline
 $A_3, A_4, A_5$ & $1.616$ & $0.106$ & $Not Reject H_0$ \\
\hline
 $A_2, A_4, A_5$ & $6.462$ & $0.000$ & $Reject H_0$ \\
\hline
 $A_2, A_3, A_5$ & $3.974$ & $0.000$ & $Reject H_0$ \\
\hline
 $A_2, A_3, A_4$ & $4.350$ & $0.000$ & $Reject H_0$ \\
\hline
 $A_1, A_4, A_5$ & $3.634$ & $0.000$ & $Reject H_0$ \\
\hline
 $A_1, A_3, A_5$ & $3.512$ & $0.000$ & $Reject H_0$ \\
\hline
 $A_1, A_3, A_4$ & $3.316$ & $0.001$ & $Reject H_0$ \\
\hline
 $A_1, A_2, A_5$ & $5.488$ & $0.000$ & $Reject H_0$ \\
\hline
 $A_1, A_2, A_4$ & $4.823$ & $0.000$ & $Reject H_0$ \\
\hline
 $A_1, A_2, A_3$ & $4.214$ & $0.000$ & $Reject H_0$ \\
\hline
\hline
 $A_1, A_2, A_3, A_4$ & $4.973$ & $0.000$ & $Reject H_0$ \\
\hline
 $A_1, A_2, A_3, A_5$ & $5.207$ & $0.000$ & $Reject H_0$ \\
\hline
 $A_1, A_2, A_4, A_5$ & $6.410$ & $0.000$ & $Reject H_0$ \\
\hline
 $A_1, A_3, A_4, A_5$ & $4.304$ & $0.000$ & $Reject H_0$ \\
\hline
 $A_2, A_3, A_4, A_5$ & $5.889$ & $0.000$ & $Reject H_0$ \\
\hline
\hline
 $A_1, A_2, A_3, A_4, A_5$ & $5.921$ & $0.000$ & $Reject H_0$ \\
\hline

\end{tabular}
\caption{Calculated test statistics  $D_n$ and p-values } \label{redata3}
\end{table}
%%%%%%%%%%%%%%%%%%%%%%%%%%%%%%%%%%%%%%%%%%%%%%%%%%%%%%%%%%%

\newpage

\section{Proof of main results}\label{mProf}

In this section, we provide detailed proofs of Theorem \ref{mainthm0} and Theorem \ref{mainthm1}. For convenience, denote $\bar{A}_{l,ij}=A_{l,ij}-\mathbb{E}[A_{l,ij}]$ and $\bar{d}_{l,i}=d_{l,i}-\mathbb{E}[d_{l,i}]$.
Denote the WDDT test statistic $D_n$ as follows
\begin{equation}\label{testD}
D_n=\frac{\mathcal{T}_n}{\sigma_n},
\end{equation}
where 
\begin{equation}\label{testMTN}
\mathcal{T}_n=\sum_{l=2}^{L}\left[\sum_{i=1}^n\left(\frac{d_{1,i}}{\sqrt{P_1}}-\frac{d_{l,i}}{\sqrt{P_1}}\right)^2-\frac{d_{1}}{P_1}-\frac{d_{l}}{P_l}\right].
\end{equation}
We shall find the leading term of $D_n$. Then we prove the leading term converges in distribution to the standard normal distribution under $H_0$ and the leading term converges to infinity under $H_1$. Since the proof of Theorem \ref{mainthm1} is easier than Theorem \ref{mainthm0}, we present the proof of Theorem \ref{mainthm1} first.

\subsection{Proof of Theorem \ref{mainthm1}}
We will find the leading term of $D_n$. Firstly, we get the order of $P_l$. 
Note that $\mathbb{E}[A_{l,ij}]=\rho_lW_{l,i}W_{l,j}$. Simple algebra yields
\begin{eqnarray} \nonumber  
P_l&=&\sum_{i\neq j\neq k}A_{l,ij}A_{l,jk}\\ \nonumber
&=&\sum_{i\neq j\neq k}\rho_l^2W_{l,i}W_{l,j}^2W_{l,k}+\sum_{i\neq j\neq k}\bar{A}_{l,ij}\bar{A}_{l,jk}\\ \label{h0eq1}
&&+\sum_{i\neq j\neq k}\bar{A}_{l,ij}\rho_lW_{l,j}W_{l,k}+\sum_{i\neq j\neq k}\bar{A}_{l,jk}\rho_lW_{l,i}W_{l,j}.
\end{eqnarray}
Next we find the order of each term in (\ref{h0eq1}). The second moment of the second term in (\ref{h0eq1}) is equal to
\begin{eqnarray*}    \mathbb{E}\left[\left(\sum_{i\neq j\neq k}\bar{A}_{l,ij}\bar{A}_{l,jk}\right)^2\right]
&=&\sum_{\substack{i\neq j\neq k\\ i_1\neq j_1\neq k_1}}\mathbb{E}\left[\bar{A}_{l,ij}\bar{A}_{l,jk}\bar{A}_{l,i_1j_1}\bar{A}_{l,j_1k_1}\right].
\end{eqnarray*}
Note that $\mathbb{E}[\bar{A}_{l,ij}]=0$ and $A_{l,ij}$ and $A_{l,i_1j_1}$ are independent if $\{i,j\}\neq\{i_1,j_1\}$. If $\{i,j\}\neq\{i_1,j_1\}$ and $\{i,j\}\neq\{k_1,j_1\}$, then
\[\mathbb{E}\left[\bar{A}_{l,ij}\bar{A}_{l,jk}\bar{A}_{l,i_1j_1}\bar{A}_{l,j_1k_1}\right]=\mathbb{E}\left[\bar{A}_{l,ij}\right]\mathbb{E}\left[\bar{A}_{l,jk}\bar{A}_{l,i_1j_1}\bar{A}_{l,j_1k_1}\right]=0.\]
Hence $\{i,j\}=\{i_1,j_1\}$ or $\{i,j\}=\{k_1,j_1\}$. Similarly, $\{j,k\}=\{i_1,j_1\}$ or $\{j,k\}=\{k_1,j_1\}$. Then $i=i_1,j=j_1,k=k_1$ or $i=k_1,j=j_1,k=i_1$. In this case,
\begin{eqnarray} \nonumber   \mathbb{E}\left[\left(\sum_{i\neq j\neq k}\bar{A}_{l,ij}\bar{A}_{l,jk}\right)^2\right]
&=&\sum_{i\neq j\neq k}\mathbb{E}\left[\big(\bar{A}_{l,ij}\bar{A}_{l,jk}\big)^2\right]+\sum_{i\neq j\neq k}\mathbb{E}\left[\big(\bar{A}_{l,jk}\bar{A}_{l,ij}\big)^2\right]\\ \nonumber
&=&2\sum_{i\neq j\neq k}\rho_l^2W_{l,i}W_{l,j}^2W_{l,k}(1-\rho_l^2W_{l,i}W_{l,j}^2W_{l,k}\big)\\ \label{evv}
&=&2\rho_l^2||W_l||_1^2(1+o(1)).
\end{eqnarray}
Similarly, we get
\begin{eqnarray}\label{2theq1}
\mathbb{E}\left[\left(\sum_{i\neq j}\bar{A}_{l,ij}\rho_lW_{l,j}W_{l,k}\right)^2\right]=O\left(\rho_l^3||W_l||_1^3||W_l||_3^3+\rho_l^3||W_l||_1^2\right).
\end{eqnarray}
Since $||W_l||_4=o(1)$ and $||W_l||_2=1$, then $||W_l||_3^3\leq \sqrt{||W_l||_4^4||W_l||_2^2}=o(1)$.
By Markov's inequality, we get
\begin{eqnarray} \label{plorder}
P_l&=&(1+o_P(1))\sum_{i\neq j\neq k}\rho_l^2W_{l,i}W_{l,j}^2W_{l,k}= (1+o_P(1))\rho_l^2||W_l||_1^2.
\end{eqnarray}

By (\ref{varsigma}), (\ref{plorder}) and the assumption that $L$ is a fixed positive integer, we have 
\begin{eqnarray} \label{sigmaorder}
\sigma_n^2&=&(1+o_P(1))\left(\frac{2(L-1)^2}{\rho_1^2||W_1||_1^2}+\sum_{l=2}^L\frac{2}{\rho_l^2||W_l||_1^2}+\sum_{l=2}^L\frac{4}{\rho_1||W_1||_1\rho_l||W_l||_1}\right).
\end{eqnarray}

Next we find the leading order of $\mathcal{T}_n$ defined in (\ref{testMTN}).
Simple calculation yields
\begin{eqnarray}\nonumber
&&\sum_{i=1}^n\left(\frac{d_{1,i}}{\sqrt{P_1}}-\frac{d_{l,i}}{\sqrt{P_l}}\right)^2-\frac{d_{1}}{P_1}-\frac{d_{l}}{P_l}\\ \nonumber
&=&\sum_{i=1}^n\left(\frac{d_{1,i}-\mathbb{E}[d_{1,i}]}{\sqrt{P_1}}+\frac{\mathbb{E}[d_{1,i}]}{\sqrt{P_1}}-\frac{\mathbb{E}[d_{l,i}]}{\sqrt{P_l}}-\frac{d_{l,i}-\mathbb{E}[d_{l,i}]}{\sqrt{P_l}}\right)^2-\frac{d_{1}}{P_1}-\frac{d_{l}}{P_l}\\ \nonumber
&=&\left(\sum_{i=1}^n\frac{\bar{d}_{1,i}^2}{P_1}-\frac{d_{1}}{P_1}\right)+\left(\sum_{i=1}^n\frac{\bar{d}_{l,i}^2}{P_l}-\frac{d_{l}}{P_l}\right)+\sum_{i=1}^n\left(\frac{\mathbb{E}[d_{1,i}]}{\sqrt{P_1}}-\frac{\mathbb{E}[d_{l,i}]}{\sqrt{P_l}}\right)^2\\ \label{coreq1}
&&-2\sum_{i=1}^n\frac{\bar{d}_{1,i}\bar{d}_{l,i}}{\sqrt{P_1P_l}}+2\sum_{i=1}^n\frac{\bar{d}_{1,i}}{\sqrt{P_1}}\left(\frac{\mathbb{E}[d_{1,i}]}{\sqrt{P_1}}-\frac{\mathbb{E}[d_{l,i}]}{\sqrt{P_l}}\right)-2\sum_{i=1}^n\frac{\bar{d}_{l,i}}{\sqrt{P_l}}\left(\frac{\mathbb{E}[d_{1,i}]}{\sqrt{P_1}}-\frac{\mathbb{E}[d_{l,i}]}{\sqrt{P_l}}\right).
\end{eqnarray}
We shall find the order of each term in (\ref{coreq1}).

Firstly, we consider the third term of (\ref{coreq1}). Note that $||W_l||_2=1$ by assumption and $\mathbb{E}[d_{l,i}]=\rho_lW_{l,i}||W_l||_1$. Direct calculation yields
\begin{eqnarray}  \nonumber
&&\sum_{i=1}^n\left(\frac{\mathbb{E}[d_{1,i}]}{\sqrt{P_1}}-\frac{\mathbb{E}[d_{l,i}]}{\sqrt{P_l}}\right)^2\\ \nonumber
&=&\sum_{i=1}^n\left(\frac{\rho_1^2W_{1,i}^2||W_1||_1^2}{P_1}+\frac{\rho_l^2W_{l,i}^2||W_l||_1^2}{P_l}-2\frac{\rho_1\rho_lW_{1,i}W_{l,i}||W_1||_1||W_l||_1}{\sqrt{P_1P_l}}\right)\\ \nonumber
&=&\frac{\rho_1^2||W_1||_1^2}{P_1}+\frac{\rho_l^2||W_l||_1^2}{P_l}-2\frac{\rho_1\rho_l||W_1||_1||W_l||_1\sum_{i=1}^nW_{1,i}W_{l,i}}{\sqrt{P_1P_l}}\\ \label{coreq2}
&=&\left(\frac{\rho_1||W_1||_1}{\sqrt{P_1}}-\frac{\rho_l||W_l||_1}{\sqrt{P_l}}\right)^2+2\frac{\rho_1\rho_l||W_1||_1||W_l||_1}{\sqrt{P_1P_l}}\left(1-\sum_{i=1}^nW_{1,i}W_{l,i}\right).
\end{eqnarray}
By the Cauchy-Schwarz inequality, we have
\begin{eqnarray*}
\sum_iW_{1;i}W_{l;i}\leq\sqrt{\Big(\sum_jW_{1;j}^2\Big)\Big(\sum_jW_{l;j}^2\Big)}=1,
\end{eqnarray*}
where equality holds if and only if $W_1=\lambda W_{l}$ for some constant $\lambda$. Since $||W_l||_2=1$ for all $l$ and $W_l\in[0,1]^n$, then $\lambda=1$. 
Hence
\begin{equation}
   1-\sum_{i=1}^nW_{1,i}W_{l,i}\geq0,
\end{equation}
and equality holds if and only if $W_1=W_{l}$.
By (\ref{plorder}), one has
\begin{equation}\label{nneweq1}
\frac{\rho_1\rho_l||W_1||_1||W_l||_1}{\sqrt{P_1P_l}}=1+o_P(1).
\end{equation}
 By (\ref{sigmaorder}), (\ref{coreq2}) and (\ref{nneweq1}), one has
\begin{eqnarray}  \nonumber
R_n:&=&\frac{1}{\sigma_n}\sum_{l=2}^L\sum_{i=1}^n\left(\frac{\mathbb{E}[d_{1,i}]}{\sqrt{P_1}}-\frac{\mathbb{E}[d_{l,i}]}{\sqrt{P_l}}\right)^2\\  \label{rnorder0} 
&\geq& \Theta\left(\frac{\sum_{l=2}^L\left(1-\sum_{i=1}^nW_{1,i}W_{l,i}\right)}{\sigma_n}\right)(1+o_P(1)).
\end{eqnarray}
Suppose $\sum_iW_{1;i}W_{l_0;i}\leq 1-\epsilon$ for some constant $\epsilon\in(0,1)$ and some $l_0\in\{2,3,\dots,L\}$. Recall that $L$ is assumed to be a fixed positive integer.
\begin{eqnarray} \label{rnorder} 
R_n&\geq&\Theta\left(\frac{1}{\sigma_n}\right)(1+o_P(1)).
\end{eqnarray}

Next we consider the first two terms of (\ref{coreq1}). 
Note that 
\begin{eqnarray}\label{thm0eq1}
\sum_{i=1}^n\frac{\bar{d}_{l,i}^2}{P_l}-\frac{d_{l}}{P_l}=\frac{1}{P_l}\sum_{i\neq j\neq k}\bar{A}_{l;ij}\bar{A}_{l,ik}+\frac{1}{P_l}\sum_{i\neq j}\bar{A}_{l;ij}^2-\frac{d_{l}}{P_l}.
\end{eqnarray}
Simple calculation yields
\begin{eqnarray*}
\sum_{i\neq j}\mathbb{E}[\bar{A}_{l;ij}^2]=\sum_{i\neq j}\rho_lW_{l,i}W_{l,j}-\sum_{i\neq j}\rho_l^2W_{l,i}^2W_{l,j}^2,
\end{eqnarray*}
and
\begin{eqnarray*}
\mathbb{E}\left[\left(\sum_{i\neq j}(\bar{A}_{l;ij}^2-\mathbb{E}[\bar{A}_{l;ij}^2])\right)^2\right]&=&\sum_{i\neq j} \mathbb{E}[(\bar{A}_{l;ij}^2-\mathbb{E}[\bar{A}_{l;ij}^2])^2]\\
&\leq&\sum_{i\neq j} \mathbb{E}[\bar{A}_{l;ij}^4]\\
&\leq& \sum_{i\neq j}\rho_lW_{l,i}W_{l,j}\\
&\leq&\rho_l||W_l||_1^2.
\end{eqnarray*}
Hence, 
\begin{eqnarray*}
\sum_{i\neq j}\bar{A}_{l;ij}^2=\sum_{i\neq j}\rho_lW_{l,i}W_{l,j}-\sum_{i\neq j}\rho_l^2W_{l,i}^2W_{l,j}^2+O_P\left(\sqrt{\rho_l||W_l||_1^2}\right).
\end{eqnarray*}
Similarly, we have
\[d_l=\sum_{i\neq j}A_{l,ij}=\sum_{i\neq j}\rho_lW_{l,i}W_{l,j}+O_P\left(\sqrt{\rho_l||W_l||_1^2}\right).\]
Then by (\ref{plorder}) and the assumption that $\rho_l=o(||W_l||_1)$, we get
\begin{eqnarray}
\frac{\sum_{i\neq j}\bar{A}_{l;ij}^2-d_l}{P_l}=\frac{O_P\left(\sqrt{\rho_l||W_l||_1^2}+\rho_l^2\right)}{P_l}=O_P\left(\frac{1}{\rho_l||W_l||_1\sqrt{\rho_l}}+\frac{\rho_l}{\rho_l||W_l||_1^2}\right).
\end{eqnarray}
Recall (\ref{evv}), (\ref{sigmaorder}), the assumption $\min_{1\leq l\leq L}\{\rho_l\}=\omega(1)$ and $\rho_l=o(||W_l||_1)$, and that $L$ is a fixed positive integer.
Hence we get
\begin{eqnarray}\label{fteq}
\frac{1}{\sigma_n}\sum_{l=2}^L\left(\sum_{i=1}^n\frac{\bar{d}_{l,i}^2}{P_l}-\frac{d_{l}}{P_l}\right)=O_P\left(\max_{1\leq l\leq L}\frac{\rho_l}{||W_l||_1}+\frac{1}{\sqrt{\min_{1\leq l\leq L}\{\rho_l\}}}\right)=o_P(1).
\end{eqnarray}

Next we consider the last two terms of (\ref{coreq1}).
Note that
\begin{eqnarray*}
\sum_{i=1}^n\frac{\bar{d}_{l,i}}{\sqrt{P_l}}\left(\frac{\mathbb{E}[d_{1,i}]}{\sqrt{P_1}}-\frac{\mathbb{E}[d_{l,i}]}{\sqrt{P_l}}\right)=\sum_{i=1}^n\frac{\bar{d}_{l,i}\mathbb{E}[d_{1,i}]}{\sqrt{P_lP_1}}-\sum_{i=1}^n\frac{\bar{d}_{l,i}\mathbb{E}[d_{l,i}]}{P_l}.
\end{eqnarray*}
It is easy to verify that
\begin{eqnarray*}
\sum_{i=1}^n\frac{\bar{d}_{l,i}\mathbb{E}[d_{l,i}]}{P_l}=\frac{\rho_l||W_l||_1}{P_l}\sum_{i\neq j}\bar{A}_{l,ij}W_{l,i}=O_P\left(\frac{\sqrt{\rho_l||W_l||_1||W_l||_3^3+\rho_l}}{\rho_l||W_l||_1}\right).
\end{eqnarray*}
 Since $||W||_4=o(1)$, then $||W||_3^3=o(1)$.
Recall (\ref{sigmaorder}), the assumption $\rho_l=o(||W_l||_1)$ and  $\min_{1\leq l\leq L}||W_l||_1=\omega(1)$,  and $L$ is a fixed positive integer. Then
\begin{eqnarray}\label{lasttweq}
\frac{1}{\sigma_n}\sum_{l=2}^L\sum_{i=1}^n\frac{\bar{d}_{l,i}\mathbb{E}[d_{l,i}]}{P_l}=o_P(R_n).
\end{eqnarray}

Now we consider the fourth term of (\ref{coreq1}). Simple calculation yields
\begin{eqnarray}\label{ftheq}
&&\frac{1}{\sqrt{P_1P_l}}\sum_i\bar{d}_{1,i}\bar{d}_{l,i}=\frac{1}{\sqrt{P_1P_l}}\sum_{i\neq j\neq k}\bar{A}_{1,ij}\bar{A}_{l,ik}+\frac{1}{\sqrt{P_1P_l}}\sum_{i\neq j}\bar{A}_{1,ij}\bar{A}_{l,ij}.
\end{eqnarray}
Note that
\begin{eqnarray*}\label{dlij}
\frac{1}{\sqrt{P_1P_l}}\sum_{i\neq j}\bar{A}_{1,ij}\bar{A}_{l,ij}=O_P\left(\frac{\sqrt{\rho_1\rho_l}\sum_{i}W_{1,i}W_{l,i}}{\rho_1||W_1||_1\rho_l||W_l||_1}\right),
\end{eqnarray*}
and
\begin{eqnarray*}
\frac{1}{\sqrt{P_1P_l}}\sum_{i\neq j\neq k}\bar{A}_{1,ij}\bar{A}_{l,ik}=O_P\left(\frac{1}{\sqrt{\rho_1||W_1||_1\rho_l||W_l||_1}}\right).
\end{eqnarray*}
Hence
\begin{eqnarray}\label{ftheq1}
&&\frac{1}{\sigma_n\sqrt{P_1P_l}}\sum_{l-2}^L\sum_i\bar{d}_{1,i}\bar{d}_{l,i}=o_P(R_n).
\end{eqnarray}

By (\ref{coreq1}), (\ref{rnorder}), (\ref{fteq}), (\ref{lasttweq}) and  (\ref{ftheq1}), the proof is complete.

\subsection{Proof of Theorem \ref{mainthm0}}

Before proving Theorem \ref{mainthm0}, we present one lemma and one proposition. For integer $t\geq3$, let 
\begin{eqnarray}\label{lemeq2}
T_{l;t}=\frac{\sqrt{2}}{\rho_l||W||_1}\sum_{i<j<k\leq t}(\bar{A}_{l,ij}\bar{A}_{l,jk}+\bar{A}_{l,ik}\bar{A}_{l,kj}+\bar{A}_{l,ji}\bar{A}_{l,ki}),
\end{eqnarray}
and
\begin{eqnarray}\nonumber
T_{1,l;t}
&=&\frac{1}{\sqrt{\rho_1\rho_{l}}||W||_1}\sum_{i<j<k\leq t}(\bar{A}_{1,ij}\bar{A}_{l,jk}+\bar{A}_{1,kj}\bar{A}_{l,ji}+\bar{A}_{1,ik}\bar{A}_{l,kj}\\ \label{lemeq1}
&&+\bar{A}_{1,jk}\bar{A}_{l,ki}+\bar{A}_{1,ij}\bar{A}_{l,ki}+\bar{A}_{1,ik}\bar{A}_{l,ij}).
\end{eqnarray}

\begin{Lemma}\label{normallem}
Under the assumption of Theorem \ref{mainthm0}, 
$T_{1;n},\dots,T_{L;n}$,$T_{1,2;n}$,$T_{1,3;n}$,$\dots,T_{1,L;n}$ jointly converges in distribution  to  the $2L-1$ dimensional standard normal distribution.
\end{Lemma}

The proof of Lemma \ref{normallem} depends on the following proposition.
   
\begin{Proposition}[\cite{HH14}]\label{martingale}
 Suppose that for every $n\in\mathbb{N}$ and $k_n\rightarrow\infty$ the random variables $X_{n,1},\dots,X_{n,k_n}$ are a martingale difference sequence relative to an arbitrary filtration $\mathcal{F}_{n,1}\subset\mathcal{F}_{n,2}$ $\subset$ $\dots$ $\subset\mathcal{F}_{n,k_n}$. If (I) $\sum_{i=1}^{k_n}\mathbb{E}(X_{n,i}^2|\mathcal{F}_{n,i-1})\rightarrow 1$ in probability,
 (II) $\sum_{i=1}^{k_n}\mathbb{E}(X_{n,i}^2I[|X_{n,i}|>\epsilon]|\mathcal{F}_{n,i-1})\rightarrow 0$ in probability for every $\epsilon>0$,
\noindent then $\sum_{i=1}^{k_n}X_{n,i}\rightarrow N(0,1)$ in distribution.
\end{Proposition}

\medskip

\noindent
{\bf Proof of Lemma \ref{normallem}:} By the Cramér–Wold theorem, we only need to prove the following sequence of random variables
\[\sum_{l=1}^L\lambda_lT_{l;n}+\sum_{l=2}^L\lambda_{1,l}T_{1,l;n}\]
converges in distribution to the standard normal distribution. Here $\lambda_l,\lambda_{1,l}$ are constants such that
\[\sum_{l=1}^L\lambda_l^2+\sum_{l=2}^L\lambda_{1,l}^2=1.\]

We employ Proposition \ref{martingale} to prove this. Under $H_0$, $W_l=W$ for all $l\in[L]$. For convenience, we let $L=3$. The proof for general $L$ is similar. Let $\lambda_1$, $\lambda_2$, $\lambda_3$, $\lambda_4$, $\lambda_5$ be fixed constants such that 
\[\lambda_1^2+\lambda_2^2+\lambda_3^2+\lambda_4^2+\lambda_5^2=1.\] 
Define
\[Y_t=\lambda_1T_{1;t}+\lambda_2T_{2;t}+\lambda_3T_{3;t}+\lambda_4T_{1,2;t}+\lambda_5T_{1,3;t},\]
for $3\leq t\leq n$, and $Y_2=0$. Then $\{Y_t\}_{t=2}^n$ is a martingale.

Let $X_t=Y_t-Y_{t-1}$ and $F_t=\{A_{l;ij},1\leq i<j\leq t,1\leq l\leq L\}$ for $t\geq3$. It is easy to verify that
\[X_t=\lambda_1R_{1;t}+\lambda_2R_{2;t}+\lambda_3R_{3;t}+\lambda_4R_{1,2;t}+\lambda_5R_{1,3;t},\]
where
\[R_{l;t}=\frac{\sqrt{2}}{\rho_l||W||_1}\sum_{i<j<t}(\bar{A}_{l,ij}\bar{A}_{l,jt}+\bar{A}_{l,it}\bar{A}_{l,tj}+\bar{A}_{l,ji}\bar{A}_{l,it}),\]
\[R_{1,l;t}=\frac{1}{\sqrt{\rho_1\rho_{l}}||W||_1}\sum_{i<j<t}(\bar{A}_{1,ij}\bar{A}_{l,jt}+\bar{A}_{1,tj}\bar{A}_{l,ji}+\bar{A}_{1,it}\bar{A}_{l,tj}+\bar{A}_{1,jt}\bar{A}_{l,ti}+\bar{A}_{1,ji}\bar{A}_{l,it}+\bar{A}_{1,ti}\bar{A}_{l,ij}).\]
Then
 $\mathbb{E}[X_{t}|F_{t-1}]=0$. Hence, $\{X_t\}_{t=3}^n$ is a martingale difference. 
 
The proof proceeds by verifying the two conditions in Proposition \ref{martingale}. We begin with verifying condition $(I)$.
It is easy to get $\mathbb{E}\big(Y_{t}|F_{t-1}\big)=Y_{t-1}$. By the property of conditional expectation, one has
\begin{eqnarray*}
\mathbb{E}\left[\sum_{t=3}^n\mathbb{E}\Big(X^2_{t}|F_{t-1}\Big)\right]&=&\mathbb{E}\left[\sum_{t=3}^n\mathbb{E}\Big((Y^2_{t}-2Y_{t}Y_{t-1}+Y^2_{t-1})|F_{t-1}\Big)\right]\\
&=&\mathbb{E}\left[\sum_{t=3}^n\mathbb{E}\big[(Y^2_{t}-Y^2_{t-1})|F_{t-1}\big]\right]=\mathbb{E}[Y^2_{n}].
\end{eqnarray*}
Note that
\[Y_n=\lambda_1T_{1;n}+\lambda_2T_{2;n}+\lambda_3T_{3;n}+\lambda_4T_{1,2;n}+\lambda_5T_{1,3;n}.\]
It is easy to check
\begin{eqnarray}\label{fineq1}
\sum_{i\neq j\neq k}\bar{A}_{l;ij}\bar{A}_{l,ik}=2\sum_{i<j<k}(\bar{A}_{l,ij}\bar{A}_{l,jk}+\bar{A}_{l,ik}\bar{A}_{l,kj}+\bar{A}_{l,ji}\bar{A}_{l,ki}),
\end{eqnarray}
and
\begin{eqnarray}\nonumber
\sum_{i\neq j\neq k}\bar{A}_{1,ij}\bar{A}_{l,ik}&=&\sum_{i<j<k}(\bar{A}_{1,ij}\bar{A}_{l,jk}+\bar{A}_{1,kj}\bar{A}_{l,ji}+\bar{A}_{1,ik}\bar{A}_{l,kj}+\bar{A}_{1,jk}\bar{A}_{l,ki}\\ \label{fineq2}
&&+\bar{A}_{1,ij}\bar{A}_{l,ik}+\bar{A}_{1,ik}\bar{A}_{l,ij}).
\end{eqnarray}
By (\ref{evv}), $\mathbb{E}[T_{l;n}^2]=1+o(1)$. Similar to (\ref{evv}), it is easy to show that $\mathbb{E}[T_{1,l;n}^2]=1+o(1)$ and $\mathbb{E}[T_{l_1;n}T_{1,l_2;n}]=0$ for $l_1\in[L]$ and $l_2\in\{2,3,\dots,L\}$. Hence $\mathbb{E}[Y_n^2]=1+o(1)$ and
\begin{eqnarray*}
\mathbb{E}\left[\sum_{t=3}^n\mathbb{E}\Big(X^2_{t}|F_{t-1}\Big)\right]=\mathbb{E}[Y^2_{n}]=1+o(1).
\end{eqnarray*}

To prove condition $(I)$ holds, we only need to show that 
\begin{equation}\label{propeq1}
\mathbb{E}\Big(\sum_{t=3}^n\mathbb{E}[X_{t}^2|F_{t-1}]\Big)^2=1+o(1).
\end{equation}
Direct calculation yields
\begin{eqnarray}\nonumber
X_t^2&=&\lambda_1^2R_{1;t}^2+\lambda_2^2R_{2;t}^2+\lambda_3^2R_{3;t}^2+\lambda_4^2R_{1,2;t}^2+\lambda_5^2R_{1,3;t}^2\\ \nonumber
&&+2\lambda_1\lambda_2R_{1;t}R_{2;t}+2\lambda_1\lambda_3R_{1;t}R_{3;t}+2\lambda_2\lambda_3R_{2;t}R_{3;t}\\ \nonumber
&&+2\lambda_1\lambda_4R_{1;t}R_{1,2;t}+2\lambda_1\lambda_5R_{1;t}R_{1,3;t}\\ \nonumber
&&+2\lambda_2\lambda_4R_{2;t}R_{1,2;t}+2\lambda_2\lambda_5R_{2;t}R_{1,3;t}\\ \nonumber
&&+2\lambda_3\lambda_4R_{3;t}R_{1,2;t}+2\lambda_3\lambda_5R_{3;t}R_{1,3;t}\\ \label{propeq2}
&&+2\lambda_4\lambda_5R_{1,2;t}R_{1,3;t}.
\end{eqnarray}
Hence we have
\begin{eqnarray}\nonumber
\sum_{t=3}^n\mathbb{E}[X_{t}^2|F_{t-1}]&=&\lambda_1^2\sum_{t=3}^n\mathbb{E}[R_{1;t}^2|F_{t-1}]+\lambda_2^2\sum_{t=3}^n\mathbb{E}[R_{2;t}^2|F_{t-1}]+\lambda_3^2\sum_{t=3}^n\mathbb{E}[R_{3;t}^2|F_{t-1}]\\ \nonumber
&&+\lambda_4^2\sum_{t=3}^n\mathbb{E}[R_{1,2;t}^2|F_{t-1}]+\lambda_5^2\sum_{t=3}^n\mathbb{E}[R_{1,3;t}^2|F_{t-1}]\\ \nonumber
&&+2\lambda_1\lambda_4\sum_{t=3}^n\mathbb{E}[R_{1;t}R_{1,2;t}|F_{t-1}]+2\lambda_2\lambda_4\sum_{t=3}^n\mathbb{E}[R_{2;t}R_{1,2;t}|F_{t-1}]\\ \nonumber
&&+2\lambda_2\lambda_5\sum_{t=3}^n\mathbb{E}[R_{1;t}R_{1,3;t}|F_{t-1}]+2\lambda_3\lambda_5\sum_{t=3}^n\mathbb{E}[R_{3;t}R_{1,3;t}|F_{t-1}]\\ \label{apropeq1}
&&+2\lambda_4\lambda_5\sum_{t=3}^n\mathbb{E}[R_{1,2;t}R_{1,3;t}|F_{t-1}].
\end{eqnarray}

Next we find the conditional expectations in (\ref{apropeq1}). Consider $R_{l;t}$ first.
\begin{eqnarray*}
\mathbb{E}\big[R_{l;t}^2|F_{t-1}\big]&=&\frac{2}{\rho_l^2||W||_1^2}\sum_{i<j<t,i_1<j_1<t}\mathbb{E}\big[(\bar{A}_{l,ij}\bar{A}_{l,jt}+\bar{A}_{l,it}\bar{A}_{l,tj}+\bar{A}_{l,ji}\bar{A}_{l,ti})\\
&&\times(\bar{A}_{l,i_1j_1}\bar{A}_{l,j_1t}+\bar{A}_{l,i_1t}\bar{A}_{l,tj_1}+\bar{A}_{l,j_1i_1}\bar{A}_{l,ti_1})|F_{t-1}\big]\\
&=&\frac{2}{\rho_l^2||W||_1^2}\sum_{i<j<t,i_1<j_1<t}\Bigg(\mathbb{E}\big[\bar{A}_{l,ij}\bar{A}_{l,jt}\bar{A}_{l,i_1j_1}\bar{A}_{l,j_1t}|F_{t-1}\big]+\mathbb{E}\big[\bar{A}_{l,ij}\bar{A}_{l,jt}\bar{A}_{l,i_1t}\bar{A}_{l,tj_1}|F_{t-1}\big]\\
&&+\mathbb{E}\big[\bar{A}_{l,ij}\bar{A}_{l,jt}\bar{A}_{l,j_1i_1}\bar{A}_{l,ti_1}|F_{t-1}\big]+\mathbb{E}\big[\bar{A}_{l,it}\bar{A}_{l,tj}\bar{A}_{l,i_1j_1}\bar{A}_{l,j_1t}|F_{t-1}\big]\\
&&+\mathbb{E}\big[\bar{A}_{l,it}\bar{A}_{l,tj}\bar{A}_{l,i_1t}\bar{A}_{l,tj_1}|F_{t-1}\big]+\mathbb{E}\big[\bar{A}_{l,it}\bar{A}_{l,tj}\bar{A}_{l,j_1i_1}\bar{A}_{l,ti_1}|F_{t-1}\big]\\
&&+\mathbb{E}\big[\bar{A}_{l,ji}\bar{A}_{l,ti}\bar{A}_{l,i_1j_1}\bar{A}_{l,j_1t}|F_{t-1}\big]+\mathbb{E}\big[\bar{A}_{l,ji}\bar{A}_{l,ti}\bar{A}_{l,i_1t}\bar{A}_{l,tj_1}|F_{t-1}\big]\\
&&+\mathbb{E}\big[\bar{A}_{l,ji}\bar{A}_{l,ti}\bar{A}_{l,j_1i_1}\bar{A}_{l,ti_1}|F_{t-1}\big]\Bigg).
\end{eqnarray*}
Note that $\bar{A}_{l,ij}$ and $\bar{A}_{l,i_1j_1}$ are independent if $\{i,j\}\neq \{i_1,j_1\}$. For $i<j<t$ and $i_1<j_1<t$,  we have
\begin{eqnarray*}
\mathbb{E}\big[\bar{A}_{l,ij}\bar{A}_{l,jt}\bar{A}_{l,i_1j_1}\bar{A}_{l,j_1t}|F_{t-1}\big]=\bar{A}_{l,ij}\bar{A}_{l,i_1j_1}\mathbb{E}\big[\bar{A}_{l,jt}\bar{A}_{l,j_1t}\big].
\end{eqnarray*}
If $j\neq j_1$, then $\mathbb{E}\big[\bar{A}_{l,jt}\bar{A}_{l,j_1t}\big]=0$. If $j=j_1$, then 
\[\mathbb{E}\big[\bar{A}_{l,jt}\bar{A}_{l,j_1t}\big]=\mathbb{E}\big[\bar{A}_{l,jt}^2\big]=(1+o(1))\rho_lW_jW_t.\]
Hence we get
\begin{eqnarray*}\label{feq}
\sum_{i<j<t,i_1<j_1<t}\mathbb{E}\big[\bar{A}_{l,ij}\bar{A}_{l,jt}\bar{A}_{l,i_1j_1}\bar{A}_{l,j_1t}|F_{t-1}\big]=(1+o(1))\sum_{i<j<t,i_1<j<t}\bar{A}_{l,ij}\bar{A}_{l,i_1j} \rho_lW_jW_t.
\end{eqnarray*}

Moreover, for $i<j<t$ and $i_1<j_1<t$, $|\{j,i_1,j_1\}|\geq2$. Then
\begin{eqnarray*}
\mathbb{E}\big[\bar{A}_{l,ij}\bar{A}_{l,jt}\bar{A}_{l,i_1t}\bar{A}_{l,tj_1}|F_{t-1}\big]=\bar{A}_{l,ij}\mathbb{E}\big[\bar{A}_{l,jt}\bar{A}_{l,i_1t}\bar{A}_{l,tj_1}\big]=0.
\end{eqnarray*}

Hence, we get
\begin{eqnarray*}
\mathbb{E}\big[R_{l;t}^2|F_{t-1}\big]&=&\frac{2(1+o(1))}{\rho_l^2||W||_1^2}\Bigg(\sum_{i<j<t,i_1<j<t}\bar{A}_{l,ij}\bar{A}_{l,i_1j} \rho_lW_jW_t+\sum_{i<j<t}\rho_l^2W_iW_t^2W_j\\
&&+\sum_{i<j<t,j<j_1<t}\bar{A}_{l,ij}\bar{A}_{l,jj_1} \rho_lW_jW_t+\sum_{i<j<t,i_1<i<t}\bar{A}_{l,ij}\bar{A}_{l,i_1i} \rho_lW_iW_t\\
&&+\sum_{i<j<t,i<j_1<t}\bar{A}_{l,ij}\bar{A}_{l,j_1i} \rho_lW_iW_t\Bigg).
\end{eqnarray*}

Similarly, it is easy to have
\begin{eqnarray*}
\mathbb{E}\big[R_{1,l;t}^2|F_{t-1}\big]&=&\frac{1}{\rho_1\rho_{l}||W||_1^2}\sum_{\substack{i<j<t\\ i_1<j_1<t}}\mathbb{E}\big[(\bar{A}_{1,ij}\bar{A}_{l,jt}+\bar{A}_{1,tj}\bar{A}_{l,ji}+\bar{A}_{1,it}\bar{A}_{l,tj}\\
&&+\bar{A}_{1,jt}\bar{A}_{l,ti}+\bar{A}_{1,ji}\bar{A}_{l,it}+\bar{A}_{1,ti}\bar{A}_{l,ij})\\
&\times&(\bar{A}_{1,i_1j_1}\bar{A}_{l,j_1t}+\bar{A}_{1,tj_1}\bar{A}_{l,j_1i_1}+\bar{A}_{1,i_1t}\bar{A}_{l,tj_1}\\
&&+\bar{A}_{1,j_1t}\bar{A}_{l,ti_1}+\bar{A}_{1,j_1i_1}\bar{A}_{l,i_1t}+\bar{A}_{1,ti_1}\bar{A}_{l,i_1j_1})|F_{t-1}\big]\\
&=&\frac{1}{\rho_1\rho_{l}||W||_1^2}\Bigg(\sum_{\substack{i<j<t\\ i_1<j<t}}\bar{A}_{1,ij}\bar{A}_{1,i_1j}\rho_{l}W_jW_t+\sum_{\substack{i<j<t\\ j<j_1<t}}\bar{A}_{1,ij}\bar{A}_{1,j_1j}\rho_{l}W_jW_t\\
&&+\sum_{\substack{i<j<t\\ i_1<j<t}}\bar{A}_{l,ji_1}\bar{A}_{l,ji}\rho_1W_jW_t+\sum_{\substack{i<j<t\\ j<j_1<t}}\bar{A}_{l,j_1j}\bar{A}_{l,ji}\rho_1W_jW_t\\
&&+\sum_{\substack{i<j<t}}\rho_1\rho_{l}W_iW_t^2W_j+\sum_{\substack{i<j<t}}\rho_1\rho_{l}W_iW_t^2W_j\\
&&+\sum_{\substack{i<j<t\\ i_1<i<t}}\bar{A}_{1,ij}\bar{A}_{1,i_1i}\rho_{l}W_iW_t+\sum_{\substack{i<j<t\\ i<j_1<t}}\bar{A}_{1,ij}\bar{A}_{1,j_1i}\rho_{l}W_iW_t\\
&&+\sum_{\substack{i<j<t\\ i_1<i<t}}\bar{A}_{l,ii_1}\bar{A}_{l,ji}\rho_1W_iW_t+\sum_{\substack{i<j<t\\ i<j_1<t}}\bar{A}_{l,j_1i}\bar{A}_{l,ji}\rho_1W_iW_t\Bigg).
\end{eqnarray*}

Note that $A_l (1\leq l\leq L)$ are independent. Then
\begin{eqnarray*}
\mathbb{E}\big[2\lambda_1\lambda_2R_{1;t}R_{2;t}+2\lambda_1\lambda_3R_{1;t}R_{3;t}+2\lambda_1\lambda_2R_{2;t}R_{3;t}|F_{t-1}\big]=0,
\end{eqnarray*}
and
\begin{eqnarray*}
\mathbb{E}\big[\lambda_1\lambda_5R_{2;t}R_{1,3;t}|F_{t-1}\big]=\mathbb{E}\big[\lambda_3\lambda_4R_{3;t}R_{1,2;t}|F_{t-1}\big]0.
\end{eqnarray*}

Note that
\begin{eqnarray*}
&&\mathbb{E}\big[R_{1;t}R_{1,l;t}|F_{t-1}\big]\\
&=&\frac{\sqrt{2}}{\rho_1\sqrt{\rho_1\rho_{l}}||W||_1^2}\sum_{\substack{i<j<t\\ i_1<j_1<t}}\mathbb{E}\big[(\bar{A}_{1,ij}\bar{A}_{1,jt}+\bar{A}_{1,it}\bar{A}_{1,tj}+\bar{A}_{1,ji}\bar{A}_{1,it})(\bar{A}_{1,i_1j_1}\bar{A}_{l,j_1t}+\bar{A}_{1,tj_1}\bar{A}_{l,j_1i_1}\\
&&+\bar{A}_{1,i_1t}\bar{A}_{l,tj_1}+\bar{A}_{1,j_1t}\bar{A}_{l,ti_1}+\bar{A}_{1,j_1i_1}\bar{A}_{l,i_1t}+\bar{A}_{1,ti_1}\bar{A}_{l,i_1j_1})|F_{t-1}\big]\\
&=&\frac{\sqrt{2}}{\rho_1\sqrt{\rho_1\rho_{l}}||W||_1^2}\Bigg(\sum_{\substack{i<j<t\\ i_1<j<t}}\bar{A}_{1,ij}\bar{A}_{l,ji_1}\rho_1W_tW_j+\sum_{\substack{i<j<t\\ j<j_1<t}}\bar{A}_{1,ij}\bar{A}_{l,jj_1}\rho_1W_tW_j\\
&&+\sum_{\substack{i<j<t\\ i_1<i<t}}\bar{A}_{1,ij}\bar{A}_{l,ii_1}\rho_1W_tW_i+\sum_{\substack{i<j<t\\ i<j_1<t}}\bar{A}_{1,ij}\bar{A}_{l,ij_1}\rho_1W_tW_i\Bigg),
\end{eqnarray*}
and
\begin{eqnarray*}
&&\mathbb{E}\big[R_{l;t}R_{1,l;t}|F_{t-1}\big]\\
&=&\frac{\sqrt{2}}{\rho_{l}\sqrt{\rho_1\rho_{l}}||W||_1^2}\sum_{\substack{i<j<t\\ i_1<j_1<t}}\mathbb{E}\big[(\bar{A}_{l,ij}\bar{A}_{l,jt}+\bar{A}_{l,it}\bar{A}_{l,tj}+\bar{A}_{l,ji}\bar{A}_{l,it})(\bar{A}_{1,i_1j_1}\bar{A}_{l,j_1t}\\
&&+\bar{A}_{1,tj_1}\bar{A}_{l,j_1i_1}+\bar{A}_{1,i_1t}\bar{A}_{l,tj_1}+\bar{A}_{1,j_1t}\bar{A}_{l,ti_1}+\bar{A}_{1,j_1i_1}\bar{A}_{l,i_1t}+\bar{A}_{1,ti_1}\bar{A}_{l,i_1j_1})|F_{t-1}\big]\\
&=&\frac{\sqrt{2}}{\rho_{l}\sqrt{\rho_1\rho_{l}}||W||_1^2}\Bigg(\sum_{\substack{i<j<t\\ i_1<j<t}}\bar{A}_{l,ij}\bar{A}_{1,ji_1}\rho_{l}W_tW_j+\sum_{\substack{i<j<t\\ j<j_1<t}}\bar{A}_{l,ij}\bar{A}_{1,jj_1}\rho_{l}W_tW_j\\
&&+\sum_{\substack{i<j<t\\ i_1<i<t}}\bar{A}_{l,ij}\bar{A}_{1,ii_1}\rho_{l}W_tW_i+\sum_{\substack{i<j<t\\ i<j_1<t}}\bar{A}_{l,ij}\bar{A}_{1,ij_1}\rho_{l}W_tW_i\Bigg),
\end{eqnarray*}
and
\begin{eqnarray*}
&&\mathbb{E}[R_{l-1,l;t}R_{l,l+1;t}|F_{t-1}]
=\frac{1}{\sqrt{\rho_{l-1}\rho_l^2\rho_{l+1}}||W||_1^2}\\
&&\times\sum_{\substack{i<j<t\\ i_1<j_1<t}}(\bar{A}_{l,ij}\bar{A}_{l+1,jt}+\bar{A}_{l,tj}\bar{A}_{l+1,ji}+\bar{A}_{l,it}\bar{A}_{l+1,tj}+\bar{A}_{l,jt}\bar{A}_{l+1,ti}+\bar{A}_{l,ji}\bar{A}_{l+1,it}+\bar{A}_{l,ti}\bar{A}_{l+1,ij})\\
&&\times (\bar{A}_{l-1,i_1j_1}\bar{A}_{l,j_1t}+\bar{A}_{l-1,tj_1}\bar{A}_{l,j_1i_1}+\bar{A}_{l-1,i_1t}\bar{A}_{l,tj_1}+\bar{A}_{l-1,j_1t}\bar{A}_{l,ti_1}+\bar{A}_{l-1,j_1i_1}\bar{A}_{l,i_1t}+\bar{A}_{l-1,ti_1}\bar{A}_{l,i_1j_1})\\
&=&\frac{1}{\sqrt{\rho_{l-1}\rho_l^2\rho_{l+1}}||W||_1^2}\Bigg(\sum_{\substack{i<j<t\\ i_1<j<t}}\bar{A}_{l-1,i_1j}\bar{A}_{l+1,ij}\rho_lW_tW_j+\sum_{\substack{i<j<t\\ j<j_1<t}}\bar{A}_{l-1,j_1j}\bar{A}_{l+1,ij}\rho_lW_tW_j\\
&&+\sum_{\substack{i<j<t\\ i_1<i<t}}\bar{A}_{l-1,i_1i}\bar{A}_{l+1,ij}\rho_lW_tW_i+\sum_{\substack{i<j<t\\ i<j_1<t}}\bar{A}_{l-1,j_1i}\bar{A}_{l+1,ij}\rho_lW_tW_i\Bigg).
\end{eqnarray*}

Now we find the second moment of each term in (\ref{apropeq1}). Firstly, we consider the first three terms of (\ref{apropeq1}).
\begin{eqnarray*}
\sum_{t=3}^n\mathbb{E}\big[R_{l;t}^2|F_{t-1}\big]&=&\frac{2}{\rho_l^2||W||_1^2}\Bigg(\sum_{i<j<t\leq n,i_1\neq i}\bar{A}_{l,ij}\bar{A}_{l,i_1j} \rho_lW_jW_t+\sum_{i<j<t\leq n}\bar{A}_{l,ij}^2\rho_lW_jW_t\\
&&+\sum_{i<j<t\leq n}\rho_l^2W_iW_t^2W_j+\sum_{i<j<t\leq n,j<j_1<t\leq n}\bar{A}_{l,ij}\bar{A}_{l,jj_1} \rho_lW_jW_t\\
&&+\sum_{i<j<t\leq n,i_1<i<t\leq n}\bar{A}_{l,ij}\bar{A}_{l,i_1i} \rho_lW_iW_t+\sum_{i<j<t\leq n,j_1\neq j}\bar{A}_{l,ij}\bar{A}_{l,j_1i} \rho_lW_iW_t\Bigg)\\
&&+\sum_{i<j<t\leq n}\bar{A}_{l,ij}^2\rho_lW_iW_t\Bigg).
\end{eqnarray*}

Since $||W||_4=o(1)$ by assumption, then
\begin{eqnarray*}
&&\mathbb{E}\left[\left(\frac{2}{\rho_l^2||W||_1^2}\sum_{i<j<t\leq n,i_1\neq i}\bar{A}_{l,ij}\bar{A}_{l,i_1j} \rho_lW_jW_t\right)^2\right]\\
&=&\frac{4}{\rho_l^4||W||_1^4}\sum_{\substack{i<j<t\leq n,i_1\neq i\\ i_2<j_2<t_2\leq n,i_3\neq i_2}}\mathbb{E}\left[\bar{A}_{l,ij}\bar{A}_{l,i_1j} \bar{A}_{l,i_2j_2}\bar{A}_{l,i_3j_2}\rho_lW_jW_t\rho_lW_{j_2}W_{t_2} \right]\\
&=&\frac{4}{\rho_l^4||W||_1^4}\sum_{\substack{i<j<t\leq n,i_1\neq i\\ t_2}}\mathbb{E}\left[\bar{A}_{l,ij}^2\bar{A}_{l,i_1j}^2\rho_lW_j^2W_t\rho_lW_{t_2} \right]\\
&\leq&\frac{4\rho_l^4||W||_1^4||W||_4^4}{\rho_l^4||W||_1^4}\\
&=&o(1).
\end{eqnarray*}

Similarly, we have

\begin{eqnarray*}
\mathbb{E}\left[\left(\frac{2}{\rho_l^2||W||_1^2}\sum_{i<j<t\leq n,j<j_1<t\leq n}\bar{A}_{l,ij}\bar{A}_{l,jj_1} \rho_lW_jW_t\right)^2\right]&=&o(1).
\end{eqnarray*}

\begin{eqnarray*}
\mathbb{E}\left[\left(\frac{2}{\rho_l^2||W||_1^2}\sum_{i<j<t\leq n,i_1<i<t\leq n}\bar{A}_{l,ij}\bar{A}_{l,i_1i} \rho_lW_iW_t\right)^2\right]&=&o(1).
\end{eqnarray*}

\begin{eqnarray*}
\mathbb{E}\left[\left(\frac{2}{\rho_l^2||W||_1^2}\sum_{i<j<t\leq n,j_1\neq j}\bar{A}_{l,ij}\bar{A}_{l,j_1i} \rho_lW_iW_t\Bigg)\right)^2\right]&=&o(1).
\end{eqnarray*}

By a similar argument, we have
\begin{eqnarray*}
\mathbb{E}\left[\left(\sum_{i<j<t\leq n}\bar{A}_{l,ij}^2\rho_lW_jW_t\right)^2\right]&=&\sum_{\substack{i<j<t\leq n\\ i_1<j_1<t_1\leq n}}\mathbb{E}\left[\bar{A}_{l,ij}^2\rho_lW_jW_t\bar{A}_{l,i_1j_1}^2\rho_lW_{j_1}W_{t_1}\right]\\
&=&\sum_{\substack{i<j<t\leq n\\ i_1<j_1<t_1\leq n\\ \{i,j\}\neq  \{i_1,j_1\}}}\mathbb{E}\left[\bar{A}_{l,ij}^2\rho_lW_jW_t\bar{A}_{l,i_1j_1}^2\rho_lW_{j_1}W_{t_1}\right]\\
&&+\sum_{\substack{i<j<t\leq n\\ t_1}}\mathbb{E}\left[\bar{A}_{l,ij}^4\rho_lW_j^2W_t\rho_lW_{t_1}\right]+\sum_{\substack{i<j<t\leq n\\ t_1}}\mathbb{E}\left[\bar{A}_{l,ij}^4\rho_lW_jW_t\rho_lW_{i}W_{t_1}\right]\\
&=&\left(\sum_{i<j<t\leq n}\rho_l^2W_iW_j^2W_t\right)^2-\sum_{i<j<t\leq n}\rho_l^4W_i^2W_j^4W_t^2\\
&&+\sum_{i<j<t\leq n,t_1}\rho_l^3W_iW_j^3W_tW_{t_1}+\sum_{i<j<t\leq n,t_1}\rho_l^3W_i^2W_j^2W_tW_{t_1}\\
&=&\left(\sum_{i<j<t\leq n}\rho_l^2W_iW_j^2W_t\right)^2(1+o(1)).
\end{eqnarray*}

Hence we get
\begin{eqnarray}\nonumber
&&\mathbb{E}\left[\left(\sum_{t=3}^n\mathbb{E}\big[R_{l;t}^2|F_{t-1}\big]\right)^2\right]\\ \nonumber
&=&(1+o(1))\frac{1}{\rho_l^4||W||_1^4}\left[2\sum_{i<j<t\leq n}\Big(\rho_l^2W_iW_j^2W_t+\rho_l^2W_i^2W_jW_t+\rho_l^2W_iW_jW_t^2\Big)\right]^2\\ \nonumber
&=&(1+o(1))\frac{1}{\rho_l^4||W||_1^4}\left[\sum_{i\neq j\neq t}\rho_l^2W_iW_j^2W_t\right]^2\\ \nonumber
&=&(1+o(1))\frac{\rho_l^4||W||_1^4}{\rho_l^4||W||_1^4}\\ \label{erlt21}
&=&1+o(1).
\end{eqnarray}

Similarly, we have

\begin{eqnarray*}
\mathbb{E}\left[\left(\sum_{t=3}^n\mathbb{E}\big[R_{1,l;t}^2|F_{t-1}\big]\right)^2\right]
=1+o(1),
\end{eqnarray*}
\begin{eqnarray*}
\mathbb{E}\left[\left(\sum_{t=3}^n\mathbb{E}[R_{l_1;t}^2|F_{t-1}]\right)\left(\sum_{t=3}^n\mathbb{E}[R_{l_2;t}^2|F_{t-1}]\right)\right]=1+o(1),
\end{eqnarray*}
\begin{eqnarray*}
\mathbb{E}\left[\left(\sum_{t=3}^n\mathbb{E}[R_{1;t}^2|F_{t-1}]\right)\left(\sum_{t=3}^n\mathbb{E}[R_{1,l;t}^2|F_{t-1}]\right)\right]=1+o(1),
\end{eqnarray*}
\begin{eqnarray*}
\mathbb{E}\left[\left(\sum_{t=3}^n\mathbb{E}[R_{l;t}^2|F_{t-1}]\right)\left(\sum_{t=3}^n\mathbb{E}[R_{1,l;t}^2|F_{t-1}]\right)\right]=1+o(1),
\end{eqnarray*}
\begin{eqnarray*}
\mathbb{E}\left[\left(\sum_{t=3}^n\mathbb{E}[R_{1,2;t}^2|F_{t-1}]\right)\left(\sum_{t=3}^n\mathbb{E}[R_{1,3;t}^2|F_{t-1}]\right)\right]=1+o(1).
\end{eqnarray*}

Since $||W||_4=o(1)$, then $||W||_3=o(1)$. It is easy to verify that
\begin{eqnarray*}
\mathbb{E}\left[\left(\sum_{t=3}^n\mathbb{E}\big[R_{1;t}R_{1,l;t}|F_{t-1}\big]\right)^2\right]=o(1),
\end{eqnarray*}
\begin{eqnarray*}
\mathbb{E}\left[\left(\sum_{t=3}^n\mathbb{E}\big[R_{l;t}R_{1,l;t}|F_{t-1}\big]\right)^2\right]=o(1),
\end{eqnarray*}
and
\begin{eqnarray*}
\mathbb{E}\left[\left(\sum_{t=3}^n\mathbb{E}\big[R_{1,2;t}R_{1,3;t}|F_{t-1}\big]\right)^2\right]=o(1).
\end{eqnarray*}

Hence, we get
\begin{eqnarray*}
\mathbb{E}\left[\left(\sum_{t=3}^n\mathbb{E}[X_{t}^2|F_{t-1}]\right)^2\right]&=&(\lambda_1^2+\lambda_2^2+\lambda_3^2+\lambda_4^2+\lambda_5^2)^2(1+o(1))\\
&=&1+o(1).
\end{eqnarray*}

 Now we check condition $(II)$ in Proposition \ref{martingale}. 
Let $\epsilon$ be a fixed positive constant.  By  the Cauchy-Schwarz inequality and Markov's inequality, we have
\begin{eqnarray}\nonumber
&&\mathbb{E}\Big[\sum_{t=3}^n\mathbb{E}\big[X_{t}^2I[|X_{t}|>\epsilon|F_{t-1}\big]\Big]\\    \nonumber
&\leq&\mathbb{E}\Big[\sum_{t=3}^n\sqrt{\mathbb{E}\big[X_{t}^4|F_{t-1}\big]\mathbb{P}[|X_{t}|>\epsilon|F_{t-1}\big]}\Big]\\  \nonumber
&\leq&\frac{1}{\epsilon^2}\mathbb{E}\Bigg[\sum_{t=3}^n\mathbb{E}\Big[X_t^4\Big|F_{t-1}\Big]\Bigg]\\   \label{twoeq0}
&=& \frac{1}{\epsilon^2}\mathbb{E}\Bigg[\sum_{t=3}^nX_t^4\Bigg].
\end{eqnarray}
Note that
\begin{equation}\label{lemxt4eq1}
X_t^4\leq 5^4\left(\lambda_1^4R_{1;t}^4+\lambda_2^4R_{2;t}^4+\lambda_3^4R_{3;t}^4+\lambda_4^4R_{1,2;t}^4+\lambda_5^4R_{1,3;t}^4\right).
\end{equation}
Recall that
\begin{eqnarray*}\label{twoeq1}
R_{l;t}=\frac{\sqrt{2}}{\rho_l||W||_1}\sum_{i<j<t}\bar{A}_{l,ij}\bar{A}_{l,jt}+\frac{\sqrt{2}}{\rho_l||W||_1}\sum_{i<j<t}\bar{A}_{l,it}\bar{A}_{l,tj}+\frac{\sqrt{2}}{\rho_l||W||_1}\sum_{i<j<t}\bar{A}_{l,ji}\bar{A}_{l,it}.
\end{eqnarray*}
It is easy to verify that
\begin{eqnarray}\nonumber
R_{l;t}^4&\leq& 3^4\Bigg[\left(\frac{\sqrt{2}}{\rho_l||W||_1}\sum_{i<j<t}\bar{A}_{l,ij}\bar{A}_{l,jt}\right)^4+\left(\frac{\sqrt{2}}{\rho_l||W||_1}\sum_{i<j<t}\bar{A}_{l,it}\bar{A}_{l,tj}\right)^4\\ \label{twoeq1}
&&+\left(\frac{\sqrt{2}}{\rho_l||W||_1}\sum_{i<j<t}\bar{A}_{l,ji}\bar{A}_{l,it}\right)^4\Bigg].
\end{eqnarray}

Direct calculation yields
\begin{eqnarray}\nonumber
&&\mathbb{E}\left[\sum_{t=3}^n\left(\frac{\sqrt{2}}{\rho_l||W||_1}\sum_{i<j<t}\bar{A}_{l,ij}\bar{A}_{l,jt}\right)^4\right]\\ \label{twoeq2}
&=&\frac{4}{\rho_l^4||W||_1^4}\sum_{\substack{1\leq i<j<t\leq n\\1\leq i_1<j_1<t\leq n\\1\leq i_2<j_2<t\leq n\\1\leq i_3<j_3<t\leq n}}\mathbb{E}\left[\bar{A}_{l,ij}\bar{A}_{l,jt}\bar{A}_{l,i_1j_1}\bar{A}_{l,j_1t}\bar{A}_{l,i_2j_2}\bar{A}_{l,j_2t}\bar{A}_{l,i_3j_3}\bar{A}_{l,j_3t}\right].
\end{eqnarray}
If $\{i,j\}\neq\{i_m,j_m\}$ for all $m=1,2,3$, then 
\begin{eqnarray*}
&&\mathbb{E}\left[\bar{A}_{l,ij}\bar{A}_{l,jt}\bar{A}_{l,i_1j_1}\bar{A}_{l,j_1t}\bar{A}_{l,i_2j_2}\bar{A}_{l,j_2t}\bar{A}_{l,i_3j_3}\bar{A}_{l,j_3t}\right]\\
&=&\mathbb{E}\left[\bar{A}_{l,ij}\right]\mathbb{E}\left[\bar{A}_{l,jt}\bar{A}_{l,i_1j_1}\bar{A}_{l,j_1t}\bar{A}_{l,i_2j_2}\bar{A}_{l,j_2t}\bar{A}_{l,i_3j_3}\bar{A}_{l,j_3t}\right]\\
&=&0.
\end{eqnarray*}
For the expectation in (\ref{twoeq2}) to be non-zero, $\{i,j\}=\{i_{m_0},j_{m_0}\}$ for some $m_0\in\{1,2,3\}$. Since $i<j$ and $i_m<j_m$, then $i=i_{m_0}$ and $j=j_{m_0}$. Without loss of generality, assume $m_0=1$. By a similar argument, we get $i_2=i_3$ and $j_2=j_3$. If $\{i,j\}=\{i_2,j_2\}$, then
\begin{eqnarray}\nonumber
&&\mathbb{E}\left[\bar{A}_{l,ij}\bar{A}_{l,jt}\bar{A}_{l,i_1j_1}\bar{A}_{l,j_1t}\bar{A}_{l,i_2j_2}\bar{A}_{l,j_2t}\bar{A}_{l,i_3j_3}\bar{A}_{l,j_3t}\right]\\ \nonumber
&=&\mathbb{E}\left[\bar{A}_{l,ij}^2\bar{A}_{l,jt}^2\bar{A}_{l,i_2j_2}^2\bar{A}_{l,j_2t}^2\right]\\ \nonumber
&=&\mathbb{E}\left[\bar{A}_{l,ij}^4\bar{A}_{l,jt}^4\right]\\ \label{twoeq3}
&\leq &\rho_l^2W_iW_j^2W_t.
\end{eqnarray}
If $j=j_2$ and $i\neq i_2$, then
\begin{eqnarray}\nonumber
&&\mathbb{E}\left[\bar{A}_{l,ij}\bar{A}_{l,jt}\bar{A}_{l,i_1j_1}\bar{A}_{l,j_1t}\bar{A}_{l,i_2j_2}\bar{A}_{l,j_2t}\bar{A}_{l,i_3j_3}\bar{A}_{l,j_3t}\right]\\ \nonumber
&=&\mathbb{E}\left[\bar{A}_{l,ij}^2\bar{A}_{l,i_2j}^2\bar{A}_{l,jt}^4\right]\\ \label{twoeq4}
&\leq&\rho_l^3W_iW_{i_2}W_j^3W_t.
\end{eqnarray}
The case $j\neq j_2$ and $i=i_2$ is similar to (\ref{twoeq4}).
If $j\neq j_2$ and $i\neq i_2$, then
\begin{eqnarray}\nonumber
&&\mathbb{E}\left[\bar{A}_{l,ij}\bar{A}_{l,jt}\bar{A}_{l,i_1j_1}\bar{A}_{l,j_1t}\bar{A}_{l,i_2j_2}\bar{A}_{l,j_2t}\bar{A}_{l,i_3j_3}\bar{A}_{l,j_3t}\right]\\ \nonumber
&=&\mathbb{E}\left[\bar{A}_{l,ij}^2\bar{A}_{l,jt}^2\bar{A}_{l,i_2j_2}^2\bar{A}_{l,j_2t}^2\right]\\ \label{twoeq5}
&\leq&\rho_l^4W_iW_j^2W_t^2W_{i_2}W_{j_2}^2.
\end{eqnarray}

By (\ref{twoeq2})-(\ref{twoeq5}) and the assumption $||W||_1=\omega(1)$ and $\rho_l=\omega(1)$, we have
\begin{eqnarray}\nonumber
\mathbb{E}\left[\sum_{t=3}^n\left(\frac{\sqrt{2}}{\rho_l||W||_1}\sum_{i<j<t}\bar{A}_{l,ij}\bar{A}_{l,jt}\right)^4\right]
&\leq&\frac{4}{\rho_l^4||W||_1^4}\left(\rho_l^4||W||_1^2+\rho_l^3||W||_1^3+\rho_l^2||W||_1^2\right)\\ \label{lem4eq3}
&=&o(1).
\end{eqnarray}

Similarly, it is easy verify that
\begin{eqnarray}\label{lem4eq4}
\mathbb{E}\left[\sum_{t=3}^n\left(\frac{\sqrt{2}}{\rho_l||W||_1}\sum_{i<j<t}\bar{A}_{l,it}\bar{A}_{l,tj}\right)^4\right]=o(1),
\end{eqnarray}
\begin{eqnarray}\label{lem4eq5}
\mathbb{E}\left[\sum_{t=3}^n\left(\frac{\sqrt{2}}{\rho_l||W||_1}\sum_{i<j<t}\bar{A}_{l,ij}\bar{A}_{l,ti}\right)^4\right]=o(1).
\end{eqnarray}

Combining (\ref{twoeq1}), (\ref{lem4eq3}), (\ref{lem4eq4}) and (\ref{lem4eq5}) yields
\begin{eqnarray}\label{twoeq7}
\mathbb{E}\left[\sum_{t=3}^nR_{l;t}^4\right]=o(1).
\end{eqnarray}

By a similar argument, one can get
\begin{eqnarray}\label{twoeq8}
\mathbb{E}\left[\sum_{t=3}^nR_{1,l;t}^4\right]=o(1).
\end{eqnarray}

Combining (\ref{twoeq0}), (\ref{lemxt4eq1}), (\ref{twoeq7}) and (\ref{twoeq8}) yields
\begin{eqnarray}\nonumber
\mathbb{E}\Big[\sum_{t=3}^n\mathbb{E}\big[X_{t}^2I[|X_{t}|>\epsilon|F_{t-1}\big]\Big]
\leq \frac{1}{\epsilon^2}\mathbb{E}\Bigg[\sum_{t=3}^nX_t^4\Bigg]
=o(1).
\end{eqnarray}
 Then the desired result follows from Proposition \ref{martingale}.

\qed

\noindent
{\bf Proof of Theorem \ref{mainthm0}:}
Now we prove Theorem \ref{mainthm0}. Under $H_0$, $W_l=W$ for all $l\in[L]$. We isolate the leading term of $D_n$ and prove the leading term convenges in distribution to the standard normal distribution.

Firstly, we prove $R_n=o_P(1)$  under $H_0$ ($R_n$ defined in (\ref{rnorder})). Since $||W_l||_2=1$ and $W_l=W$ for all $l\in[L]$, then $\sum_{i=1}^nW_{1,i}W_{l,i}=1$. By (\ref{coreq2}), we have
\begin{eqnarray*}  \nonumber
\sum_{i=1}^n\left(\frac{\mathbb{E}[d_{1,i}]}{\sqrt{P_1}}-\frac{\mathbb{E}[d_{l,i}]}{\sqrt{P_l}}\right)^2&=&\left(\frac{\rho_1||W_1||_1}{\sqrt{P_1}}-\frac{\rho_l||W_l||_1}{\sqrt{P_l}}\right)^2\\
&=&\frac{1}{P_1P_l}\frac{\left(\rho_1^2||W||_1^2P_l-\rho_l^2||W||_1^2P_1\right)^2}{\left(\rho_1||W||_1\sqrt{P_l}+\rho_l||W||_1\sqrt{P_1}\right)^2}.
\end{eqnarray*}
By (\ref{h0eq1}), we have
\begin{eqnarray} \nonumber  
&&\rho_1^2||W||_1^2P_l-\rho_l^2||W||_1^2P_1\\ \nonumber  
&=&\rho_1^2||W||_1^2\sum_{i\neq j\neq k}\bar{A}_{l,ij}\bar{A}_{l,jk}+\rho_1^2||W||_1^2\sum_{i\neq j\neq k}\bar{A}_{l,ij}\rho_lW_{j}W_{k}+\rho_1^2||W||_1^2\sum_{i\neq j\neq k}\bar{A}_{l,jk}\rho_lW_{i}W_{j}\\ \nonumber
&&-\rho_l^2||W||_1^2\sum_{i\neq j\neq k}\bar{A}_{1,ij}\bar{A}_{1,jk}-\rho_l^2||W||_1^2\sum_{i\neq j\neq k}\bar{A}_{1,ij}\rho_1W_{j}W_{k}\\ \label{h0eq2}
&&-\rho_l^2||W||_1^2\sum_{i\neq j\neq k}\bar{A}_{l,jk}\rho_1W_{i}W_{j}.
\end{eqnarray}
By (\ref{evv}) and (\ref{2theq1}), one has
\[\sum_{i\neq j\neq k}\bar{A}_{l,ij}\bar{A}_{l,jk}=O_P\left(\rho_l||W||_1\right),\]
\[\sum_{i\neq j\neq k}\bar{A}_{l,ij}\rho_lW_{j}W_{k}=O_P\left(\sqrt{\rho_l^3||W||_1^3||W||_3^3+\rho_l^2||W||_1^2}\right).\]
Since $||W||_3^3=o(1)$, then
\begin{eqnarray*}  \nonumber
\sum_{i=1}^n\left(\frac{\mathbb{E}[d_{1,i}]}{\sqrt{P_1}}-\frac{\mathbb{E}[d_{l,i}]}{\sqrt{P_l}}\right)^2
&=&o_P\left(\frac{1}{\sigma_n}\right).
\end{eqnarray*}
Hence 
\begin{equation}\label{rnoorder}
R_n=o_P(1).
\end{equation}

Next we prove the last two terms of (\ref{coreq1}) are negligible.
Direct calculation yields
\begin{eqnarray*}\nonumber
\sum_{i=1}^n\frac{\bar{d}_{l,i}}{\sqrt{P_l}}\left(\frac{\mathbb{E}[d_{1,i}]}{\sqrt{P_1}}-\frac{\mathbb{E}[d_{l,i}]}{\sqrt{P_l}}\right)&=&\sum_{i=1}^n\frac{\bar{d}_{l,i}}{\sqrt{P_l}}\frac{(\sqrt{P_l}\rho_1W_i||W||_1-\sqrt{P_1}\rho_lW_i||W||_1)}{\sqrt{P_1P_l}}\\ \nonumber
&=&\frac{(\rho_1^2P_l-\rho_l^2P_1)||W||_1}{P_l\sqrt{P_1}(\sqrt{P_l}\rho_1+\sqrt{P_1}\rho_l)}\sum_{i=1}^n\bar{d}_{l,i}W_i\\ \label{coreq7}
&=&\frac{(\rho_1^2P_l-\rho_l^2P_1)||W||_1}{P_l\sqrt{P_1}(\sqrt{P_l}\rho_1+\sqrt{P_1}\rho_l)}O_P\left(\sqrt{\rho_l||W||_1||W||_3^3+\rho_l}\right).
\end{eqnarray*}
By (\ref{plorder}) and (\ref{h0eq2}), we get
\begin{eqnarray}\label{coreq6}
\sum_{i=1}^n\frac{\bar{d}_{l,i}}{\sqrt{P_l}}\left(\frac{\mathbb{E}[d_{1,i}]}{\sqrt{P_1}}-\frac{\mathbb{E}[d_{l,i}]}{\sqrt{P_l}}\right)  
&=&o_P\left(\frac{1}{\sigma_n}\right).
\end{eqnarray}

By (\ref{coreq1}), (\ref{thm0eq1}), (\ref{fteq}), (\ref{ftheq}), (\ref{rnoorder}) and (\ref{coreq6}), we have
\begin{eqnarray}\nonumber
D_n&=&Z_n+o_P(1),
\end{eqnarray}
where
\begin{eqnarray}\nonumber
Z_n=\frac{1}{\sigma_n}\sum_{l=2}^{L}\Big(\frac{1}{P_1}\sum_{i\neq j\neq k}\bar{A}_{1,ij}\bar{A}_{1,ik}+\frac{1}{P_l}\sum_{i\neq j\neq k}\bar{A}_{l;ij}\bar{A}_{l,ik}-\frac{2}{\sqrt{P_1P_l}}\sum_{i\neq j\neq k}\bar{A}_{1,ij}\bar{A}_{l,ik}\Big).
\end{eqnarray}

By (\ref{fineq1}) and (\ref{fineq2}),
 $Z_n$ is a weighted sum of  $T_{1;n},\dots,T_{L;n},T_{1,2;n},T_{1,3;n},\dots,T_{1,L;n}$ defined in (\ref{lemeq2}) and (\ref{lemeq1}), that is,
\[Z_n=\frac{1}{\sigma_n}\left(\frac{\sqrt{2}(L-1)\rho_1||W||_1}{P_1}T_{1;n}+\sum_{l=2}^L\frac{\sqrt{2}\rho_l||W||_1}{P_l}T_{l;n}-\sum_{l=2}^L\frac{2\rho_1\rho_l||W||_1^2}{\sqrt{P_1P_l}}T_{1,l;n}\right).\]
By Lemma \ref{normallem}, the vector $(T_{1;n},\dots,T_{L;n},T_{1,2;n},T_{1,3;n},\dots,T_{1,L;n})$ jointly converges in distribution  to  the $2L-1$ dimensional standard normal distribution. By (\ref{sigmaorder}) and the Slutsky's theorem, $Z_n$ converges in distribution to the standard normal distribution. The proof is complete.

\qed

\end{document}